\documentclass{aa}

\usepackage{epsfig,graphicx,natbib}
\usepackage{comment}
\usepackage{amssymb}
\usepackage{amsmath}
\usepackage{mathrsfs}
\usepackage{color}
\usepackage{subcaption}
\usepackage{booktabs}
\usepackage{xspace}
\usepackage{float}
\usepackage{breqn}
\usepackage{algpseudocode}
\usepackage{algorithm}
\usepackage{hyperref}
\usepackage[raggedright]{sidecap}
\usepackage{listings}
\usepackage{multirow}
\usepackage{longtable}
\usepackage{ltablex}
\usepackage{minted}
\usepackage{makecell}
\usepackage{xtab}
\usepackage{pdflscape}
\usepackage{arydshln}

\usepackage{tikz}
\usetikzlibrary{arrows.meta,shapes,positioning,calc,fit,backgrounds}

\usepackage{todonotes}

\newcommand{\sgra}{Sgr~A$^*$\xspace}
\newcommand{\mes}{M87$^*$\xspace}

\newcommand{\casa}{\texttt{CASA}}

\newcommand{\almaproj}[1]{\mbox{ADS/JAO.ALMA\#{#1}}}

\newcommand{\internaleht}[1]{#1}

\newcolumntype{L}{>{\raggedright\arraybackslash}X} 

\usepackage[edges]{forest}
\forestset{
  folder/.style={
    draw,
    rectangle,
    rounded corners,
    inner sep=2pt,
    edge=dashed,
    font=\ttfamily\bfseries
  },
  file/.style={
    font=\ttfamily,
    edge=dotted
  }
}

\definecolor{foldergreen}{RGB}{95,183,54}
\newlength{\treeindent}
\newlength{\levelsep}
\defcitealias{Issaoun2022}{Issaoun, Wielgus et al. 2022}
\defcitealias{Jorstad2022}{Jorstad, Wielgus et al. 2022}
\defcitealias{Mueller2023}{Paper I}
\defcitealias{Mus2023}{Mus (PhD Thesis)}
\defcitealias{Mus2023b}{Paper II}
\defcitealias{Zhang2008}{Zhang \& Li (2007)}
\defcitealias{Pardalos2017}{Pardalos et al. 2017}
\defcitealias{EHTmemo}{Albetonsa et al. 2025}

\colorlet{punct}{red!60!black}
\definecolor{background}{HTML}{EEEEEE}
\definecolor{delim}{RGB}{20,105,176}
\colorlet{numb}{magenta!60!black}

\lstdefinelanguage{json}{
    basicstyle=\normalfont\ttfamily,
    numberstyle=\scriptsize,
    stepnumber=1,
    numbersep=8pt,
    showstringspaces=false,
    breaklines=true,
    frame=lines,
    backgroundcolor=\color{background},
    literate=
     *{0}{{{\color{numb}0}}}{1}
      {1}{{{\color{numb}1}}}{1}
      {2}{{{\color{numb}2}}}{1}
      {3}{{{\color{numb}3}}}{1}
      {4}{{{\color{numb}4}}}{1}
      {5}{{{\color{numb}5}}}{1}
      {6}{{{\color{numb}6}}}{1}
      {7}{{{\color{numb}7}}}{1}
      {8}{{{\color{numb}8}}}{1}
      {9}{{{\color{numb}9}}}{1}
      {:}{{{\color{punct}{:}}}}{1}
      {,}{{{\color{punct}{,}}}}{1}
      {\{}{{{\color{delim}{\{}}}}{1}
      {\}}{{{\color{delim}{\}}}}}{1}
      {[}{{{\color{delim}{[}}}}{1}
      {]}{{{\color{delim}{]}}}}{1},
}

\begin{document}

\title{VAPOLA - A multi-year, multi-band polarization survey of AGN and \sgra at mm wavelengths with ALMA }
\subtitle{I. Survey Overview and Science-Ready Archival Products}

\author{Alejandro Mus \inst{1,2,13,14,15}
\and Ciriaco Goddi \inst{1,2,3}\fnmsep\thanks{Corresponding author:  \email{cgoddi@gmail.com
}}
\and Douglas Carlos \inst{3,4}
\and Vincenzo Galluzzi\inst{4,12}
\and Ezequiel Albentosa-Ruiz\inst{5,6}
\and Ivan Martí-Vidal \inst{5,6}
\and Hugo Messias \inst{7,8}
\and Kazi L. J. Rygl \inst{4} 
\and Geoffrey B. Crew\inst{9}
\and Lynn D. Matthews\inst{9} 
\and Elisabetta Liuzzo \inst{4}
\and Nicola Marchili \inst{4}
\and Raphael P. Rolim \inst{3}
\and Mariafelicia De Laurentis \inst{10}
\and Rocco Lico \inst{4,11}
\and Cristiano Urban \inst{12}
}

\institute{
  Dipartimento di Fisica, Università degli Studi di Cagliari, SP Monserrato-Sestu km 0.7, I-09042 Monserrato, Italy
  \and 
  INAF - Osservatorio Astronomico di Cagliari, via della Scienza 5, I-09047 Selargius (CA), Italy
  \and
  Universidade S\~ao Paulo, Instituto de Astronomia, Geof\'isica e Ci\'eencias Atmosf\'ericas, Departamento de Astronomia, S\~ao Paulo, SP 05508-090, Brazil
  \and
  INAF-Istituto di Radioastronomia, Via P. Gobetti 101, I-40129 Bologna, Italy
  \and 
  Departament d'Astronomia i Astrofísica, Universitat de València, C. Dr. Moliner 50, E-46100 Burjassot, València, Spain
  \and
  Observatori Astronòmic, Universitat de València, C. Catedrático José Beltrán 2, E-46980 Paterna, València, Spain
  \and
  ESO Vitacura, Alonso de Córdova 3107,Vitacura, Casilla 19001, Santiago de Chile, Chile
  \and
  Joint ALMA Observatory, Alonso de Córdova 3107, Vitacura 763-0355, Santiago, Chile
\and
Massachusetts Institute of Technology Haystack Observatory, 99 Millstone Road, Westford, MA 01886, USA\label{inst9}
  \and
  Dipartimento di Fisica `E. Pancini'', Università di Napoli Federico II'', Compl. Univ. di Monte S. Angelo, Edificio G, Via Cinthia, I-80126, Napoli, Italy
  \and
  Instituto de Astrof\'isica de Andaluc\'ia-CSIC, Glorieta de la Astronom\'ia s/n, E-18008 Granada, Spain
  \and
  INAF–Italian Centre for Astronomical Archives c/o Osservatorio di Trieste, via Giambattista Tiepolo, 11, I-34131, Trieste, Italy
  \and
  Soft Computing, Image Processing and Aggregation Research Group (SCOPIA) \& Modelling and Imaging Radio Astronomical Data (MIRADA), Math and Comp. Sci. Department, University of Balearic Islands, Spain,
  \and
  Artificial Intelligence Research, Institute of the Balearic Islands (IAIB)
  \and
  Health Research Institute of the Balearic Islands (IdISBa)
}

\date {Received  / Accepted}

\authorrunning{Mus}
\titlerunning{VAPOLA, a new ALMA polarization data repository for  AGN VLBI targets}

\abstract
{The Atacama Large Millimeter/submillimeter Array (ALMA) is the most sensitive interferometric array at millimeter and submillimeter wavelengths. Through the ALMA Phasing System (APS), it can participate in global Very Long Baseline Interferometry (VLBI) arrays, significantly enhancing their sensitivity and resolution. However, processing and analyzing the ALMA  data obtained in APS mode during  VLBI observations remains a complex task, requiring specialized expertise and time-consuming calibration and imaging procedures.
}%
{
In this paper, we present \textbf{VAPOLA}—the first online, multi-epoch, multi-band repository of high-level data products from ALMA observations of active galactic nuclei (AGN) and Sgr A* during global VLBI campaigns.
}
{VAPOLA is built on an automated pipeline that processes fully calibrated ALMA (QA2) data, generating science-ready products with minimal user intervention.
}
{The repository includes fully calibrated interferometric visibilities, full-Stokes images across individual and combined spectral windows, polarimetric and spectral index maps, as well as tabulated polarimetric parameters from visibility-domain polarization fitting. 
By offering ready-to-use data through a user-friendly web portal, VAPOLA enables non-expert users to perform advanced science analyses without needing in-depth knowledge of ALMA polarization calibration and imaging procedures.
}
{This resource will facilitate a broad range of scientific investigations, including the characterization of magnetic field properties in accretion flows and relativistic jets, the structure and kinematics of dusty and molecular tori in AGN, and absorption studies of the interstellar medium toward the Galactic Center. In addition, the dataset provides source-integrated parameters and calibration metadata essential for refining VLBI calibration and imaging workflows as well as for placing robust observational constraints on theoretical models of supermassive black holes and their environments.
}

\maketitle

\section{Introduction}\label{sec:obs}

Radio interferometry is a powerful technique that combines signals from multiple antennas to emulate a much larger aperture, providing high angular resolution by exploiting phase information in the measured visibilities~\citep[e.g.,][]{
Thompson2017}. 
Arrays operating at radio to submillimeter wavelengths have enabled transformative discoveries across astrophysics, ranging from star formation and complex astrochemistry in the interstellar medium (ISM), to mapping the structure of the Milky Way and rotation curves of external galaxies, to imaging relativistic jets in active galactic nuclei (AGN), and placing key constraints on the Cosmic Dawn and the Epoch of Reionization in cosmology~\citep{Andrews2018, ASTROCHEMISTRY, Reid2019, THINGS, Lister2018, COSMOLOGY}. 

Among these, the Atacama Large Millimeter/submillimeter Array (ALMA) is unmatched in sensitivity and resolution at millimeter and submillimeter wavelengths. Beyond standalone interferometry, ALMA can be phased to act as a single large antenna in Very Long Baseline Interferometry (VLBI) arrays through the ALMA Phasing System (APS)~\citep{Matthews2018, Goddi2019, Crew2023}. In this mode, ALMA has participated in international VLBI campaigns with the Event Horizon Telescope (EHT) and the Global Millimeter VLBI Array (GMVA), providing an order-of-magnitude gain in sensitivity on the longest baselines at Bands 3 (86\,GHz), 6 (230\,GHz), and more recently, Band 7 (345\,GHz).
This capability has enabled landmark results, including the first horizon-scale images of supermassive black holes: M87* at the center of the Virgo A galaxy~\citep{eht2019a,eht2019b,eht2019c,eht2019d,eht2019e,eht2019f,eht2021a,eht2021b,eht2024m87}, and Sgr~A* at the center of the Milky Way~\citep{eht2022a,eht2022b,eht2022c,eht2022d,eht2022e,eht2022f,eht2023}. More recently, observations at 345\,GHz have pushed the VLBI frontier further in angular resolution~\citep{Crew2023, Raymond2024, Goddi2025}.

While ALMA data offer exceptional scientific potential, obtaining science-ready products
(i.e., fully calibrated visibilities and advanced imaging products) still requires
substantial manual effort. Users must download raw data and associated metadata packages
from the ALMA archive, execute the \texttt{scriptForPI}, and perform custom imaging and
analysis. These challenges are significantly exacerbated for polarization observations,
which demand highly specialized expertise, particularly when acquired in APS mode.
In this mode, standard interferometric visibilities are affected by additional instrumental
and observational complexities, resulting in data structures that differ markedly from
those of regular ALMA observations \citep[][see also Sect.~\ref{sec:qa2}]{Goddi2019}. 

Taken together, these factors substantially
complicate data handling, creating barriers for non-expert users and limiting the broader
scientific exploitation of ALMA APS-mode interferometric data obtained during VLBI campaigns.

To overcome these limitations, we introduce \textbf{VAPOLA} (\textbf{V}LBI \textbf{A}GN \textbf{POL}arization data with \textbf{A}LMA), a curated and user-friendly repository of science-ready ALMA products derived from VLBI observations since 2017 (Cycle~4), spanning Bands~3, 6, and~7. 
VAPOLA is the first publicly accessible data repository dedicated to AGN and black hole targets observed with ALMA in VLBI mode, providing full-polarization, multi-epoch, and multi-band high-level data products. 
These include calibrated visibilities, polarimetric images and tables, spectral index and Faraday rotation maps, and---for time-variable sources such as Sgr~A*---light curves and time-domain diagnostics. The repository is hosted at the Italian Center for Astronomical Archives (IA2) and is regularly updated as new observations become publicly available. By minimizing manual
intervention through a largely automated processing pipeline, VAPOLA ensures internal consistency, reduces human bias, and broadens access to ALMA polarimetric data beyond the expert community.

The importance of maintaining accessible collections of advanced data products continues to grow with the rapid increase in telescope capabilities and data volumes~\citep{Villard2025}. 
 Several major community efforts, including MOJAVE, VLBA-BU-BLAZAR, POLAMI, TELAMON, ROBIN, and TANAMI,  are providing long-term polarization and monitoring data that have
become essential for studies of AGN variability and jet physics across a wide range of frequencies and angular scales \citep{Jorstad2005, Ojha2010, Lister2018, Agudo2018a, Agudo2018b, Thum2018, Eppel2024, Marchili2025}. 
Since the start of ALMA science operations in 2011, the ALMA Compact Array (ACA) has
conducted a long-term, full-polarization monitoring program of  calibrators, known as the \textit{Grid Survey} (GS). The data are systematically processed using the
\textit{Analytic Matrix for ALMA POLArimetry} (AMAPOLA\footnote{\url{https://www.alma.cl/~skameno/AMAPOLA/}}),
delivering Stokes parameters and polarization fractions of the GS sources \citep{Kameno2023b}. These polarization measurements  are highly valuable for the planning and calibration of PI-driven science projects and
for characterizing long-term intrinsic mm/sub-mm variability. 
VAPOLA is complementary to these efforts as it provides science-ready, full-polarization ALMA products from VLBI observing campaigns, bridging a critical gap
between high-resolution VLBI studies and single-dish polarimetric observations.

This paper is structured as follows. In Sect.~\ref{sec:obs}, we summarize the VLBI campaigns that provide the observational basis for VAPOLA. Sect.~\ref{sec:qa2} outlines the QA2 process, which supplies the pipeline inputs. Sect.~\ref{sec:pipeline} details the data processing steps implemented in our pipeline. Sect.~\ref{sec:architecture} describes the structure and organization of the repository. In Sect.~\ref{sec:science_case}, we present science cases that benefit from this repository. Finally, Sect.~\ref{sec:conclusions} provides a summary and outlook.

\section{VLBI campaigns with ALMA}
\label{sec:obs}
\begin{table*}
\caption{Frequency Settings of ALMA-VLBI  observations}
\centering  
\small
\begin{tabular}{ccccccclc} 
\hline\hline                  
\noalign{\smallskip}
 Band & \multicolumn{4}{c}{ Central Freq. (GHz)} & Chan. Width & No. Spec.  & Beam size & Integ. time \\
  ($\lambda$)            &    SPW\,0  &  SPW\,1 &   SPW\,2 &  SPW\,3 &    (MHz) & Chans. & arcsec &  (s) \\
\noalign{\smallskip}
\hline
\noalign{\smallskip}  
1 (7\,mm)      &    41.17     &  43.17        &  45.17      & 47.115      &   7.8125 & 240 & 2'' - 6'' & 4.03   \\
3 (3\,mm)      &    86.268     & 88.268       &  98.328      & 100.268      &   7.8125 & 240 & 2'' - 6'' & 4.03   \\
6 (1.3\,mm)      &   213.100       & 215.100        &  227.100       &  229.100       &   7.8125  & 240 & 0.3'' - 2'' &  4.03    \\
7 (0.87\,mm)      &   335.6       & 337.541        &  347.600      &  349.600       &   7.8125  & 240 & 0.3'' - 0.6'' & 4.03    \\
\noalign{\smallskip}
\hline
\end{tabular}
\label{table:freq}  
\\
\tablefoot{ALMA spectral setups: central frequencies of the four spectral windows (SPW 0 to SPW 3) in each band, channel width, number of spectral channels, synthesized beam size range, and correlation integration time.}
\end{table*}

The APS is offered to the community for VLBI science observations in ALMA Bands 1($\lambda\approx$7 mm; $\nu\approx$44\,GHz),  3 ($\lambda\approx$3 mm; $\nu\approx$86\,GHz), 6 ($\lambda\approx$1.3 mm; $\nu\approx$230\,GHz), and 7 ($\lambda\approx$0.87 mm; $\nu\approx$343\,GHz), in coordination with the Global mm-VLBI Array (GMVA) and the Event Horizon Telescope (EHT), respectively.
Across all ALMA bands, the standard spectral setup consists of four spectral windows (SPWs), each with a bandwidth of 1875,MHz. For dual-sideband receivers (bands 3, 6, 7), two SPWs are  placed in the lower sideband and two in the upper sideband. Each SPW comprises 240 channels, yielding a spectral resolution of 7.8125,MHz.
 Table~\ref{table:freq} summarizes the frequency setups offered through ALMA Cycle~11 (2025).

The first science observations using the APS were conducted in April 2017 (Cycle 4; \citealt{2019Msngr}). Since then, phased-ALMA has participated in annual global VLBI campaigns through Cycle 11 (2025), with the exception of 2019 and 2020. ALMA data are acquired simultaneously with VLBI recordings during these campaigns.
VLBI observations are typically scheduled in two-week observing windows, usually in April, except for Band~6 in 2022, which was observed in March. During phased-array operations, the 7\,m antennas of the ACA are excluded, while 37--42 of the 12\,m antennas are phased (though this number may vary in future cycles). The phasing radius also changes between campaigns, ranging from 180\,m in 2017 to 1000\,m in 2023.
Each observing run is composed of multiple ``tracks,'' defined as continuous observing sessions lasting up to 15 hours, often overnight, though daytime sessions also occur. To optimize sensitivity and $uv$-coverage, multiple VLBI projects may be interleaved within a single track, leading to some sources being observed on multiple nights within the same campaign window.

To date, the ALMA–VLBI array has supported 2 unique PI-led projects across Band 1, 20 projects across Band 3, 40 in Band 6, and 5 in Band 7. Table~\ref{tab:list_projects} provides a sample overview of the ALMA tracks and associated projects, with relevant information about the observation.


\section{Data processing}

We adopted a two-step data processing strategy. First, we performed a full-polarization calibration of the raw data downloaded from the ALMA Science Archive for all relevant VLBI projects (Sect.~\ref{sec:qa2}). The resulting calibrated visibility data were then used as input for an automated analysis pipeline that prepares final science products and data releases (Sect.~\ref{sec:pipeline}).

{Throughout this section, ``ALMA data" refers exclusively to the standard interferometric visibilities produced by the ALMA Baseline Correlator in APS mode during VLBI campaigns. The VLBI baseband recordings and correlated VLBI visibilities themselves are outside the scope of VAPOLA}.

\subsection{Raw data full-polarization calibration}
\label{sec:qa2}
During APS operation, the signal path from the antennas to the correlator differs from that of standard ALMA interferometry \citep{Matthews2018, Goddi2019}. As a result,  
calibration must be  performed independently for VLBI and non-VLBI scans. Throughout this work, we define
\emph{VLBI scans} as visibilities acquired while ALMA was phased up and participating as a single station in the global VLBI array, whereas \emph{non-VLBI scans} refer to data obtained in standalone ALMA
interferometric mode.

The raw ALMA data were calibrated using the Common Astronomy Software
Applications (\textsc{CASA}) package \citep{CASA}, following the specialized pALMA-VLBI Quality Assurance Level~2 (QA2) procedures developed for APS-mode observations and described in \citet{Goddi2019}. In the following subsections, we describe a series of additional refinements introduced in this work to improve upon the baseline QA2 calibration strategy.

\subsubsection{Absolute flux-density scale}
\label{app:fluxscale}

As described in \citet{Goddi2019}, amplitude calibration and absolute flux-density scaling in VLBI observations with ALMA largely follow standard ALMA procedures, with two key differences. First, amplitude gains must be derived separately for VLBI and non-VLBI scans. Second, system temperatures ($T_{\rm sys}$) corrections, although measured regularly to track atmospheric opacity, are not applied during phased-array calibration in order to avoid biasing the phased sum \citep{Goddi2019}. 

While the dominant opacity effects are corrected through self-calibration, $T_{\rm sys}$ measurements can be used to account for second-order opacity differences arising from mismatched elevations between science targets and primary flux calibrators. Following \citet{Goddi2019}, we compute the opacity-corrected flux density $S^S_\tau$ for a source $S$ as:
\begin{equation}
S^S_{\tau} = \left( \frac{\left< g_a^S(t) T^S_{\rm sys}(t) \right>}{\left< g_a^{P}(t) T^P_{\rm sys}(t) \right>} \right)^2 S^S_{\rm QA2},
\label{TauFluxEq}
\end{equation}
where $g_a(t)$ is the antenna-based amplitude gain and $P$ denotes the primary flux calibrator.

In contrast to previous implementations \citep{Goddi2019, Goddi2021}, where this correction was applied post-QA2 as a scaling to tabulated flux densities, we incorporate it directly into the calibration process, immediately after gain calibration. This results in opacity-corrected amplitude gains and ensures internal consistency throughout the calibration chain (see also \citealt{Crew2023, Goddi2025}).

Absolute flux calibration follows standard ALMA practice. When available, Solar System Objects (SSOs) are used as primary calibrators. Otherwise, quasars from the ALMA Compact Array (ACA) flux monitoring program are adopted, using the \texttt{GetALMAFlux} tool within the CASA \texttt{analysisUtil} package. Calibrator selection is optimized to ensure flux consistency across all sources observed within each VLBI campaign.

The accuracy of the absolute flux-density scale in VLBI observations has been assessed in several previous studies \citep{Goddi2019, Goddi2021, Crew2023, Goddi2025} by comparing ALMA flux densities of VLBI targets with independent GS measurements from ACA monitoring programs. These works show that the flux densities derived from VLBI-mode ALMA observations are consistent with standard ALMA calibration uncertainties, namely $\sim$5\% in Band~3 and $\sim$10\% in Bands~6 and 7 \citep{Remijan2019}. We therefore adopt these values as representative uncertainties for the VAPOLA data products.

Starting from Cycle~7 (2021 campaign onward), an additional observational strategy has been introduced in which phased science targets (or their associated phasors) are also observed in non-VLBI scans with phase-calibrator intents. In these scans, the APS is inactive, allowing standard calibration procedures—including the application of $T_{\rm sys}$—to be used to derive an independent absolute flux scale. This can then be transferred to VLBI scans, potentially improving the treatment of opacity effects.  
This approach is not yet implemented in the current VAPOLA calibration scheme. Future pipeline developments (Cycles~10 and beyond) will evaluate its adoption and quantify its impact on the absolute flux scale. Updated calibration strategies will be incorporated into future VAPOLA data releases as appropriate (see Appendix~\ref{data_releases}).

\subsubsection{Polarization calibration}
\label{polcal}
Full polarization calibration is essential for science-quality VLBI analysis \citep{polconvert, Goddi2019}. ALMA’s linearly polarized feeds simultaneously record orthogonal X and Y signals. All phased-array visibilities are aligned to a reference antenna whose phase is set to zero in both polarizations, leaving residual XY phase offsets that must be corrected.

In some cases, the cross-hand phase shows steep slopes with frequency, indicating significant residual delays ($>$0.2 ns) across 2 GHz band \citep[e.g. see Fig. 9 in][]{Crew2023}. These arise because standard bandpass and gain calibration only correct parallel-hand delays, and baseband delay corrections are not applied when the APS is active \citep{Matthews2018}. While phasing corrections mitigate these effects, 
\internaleht{an X-vs-Y instrumental delay of the reference antenna remains and is shared by all visibilities}.
These are corrected using \texttt{gaincal} with \texttt{gaintype=$^\prime$KCROSS$^\prime$}, separately for VLBI and non-VLBI scans. This approach improves upon the QA2 procedure outlined in \citet{Goddi2019} and ensures accurate circular polarization conversion \citep{polconvert}.

\subsubsection{Bandpass solutions for APS scans}

{As detailed in \citet{Goddi2019}, two distinct bandpass calibration tables are employed. 
The first is derived from standard calibration using the non-VLBI scans and is applied to them within the VLBI observations. 
The second table is a duplicate of the first, but with all phase terms set to zero, and is applied to the VLBI scans. 
This distinction is required due to the intrinsic differences between VLBI scans and non-VLBI  scans. 
In ALMA mode, any discrepancy between X and Y bandpass phases reflects residual cross-delays from the ALMA correlator model, which is not used in APS mode. 
Conversely, during APS operations, the phasing process introduces additional X–Y phase terms (in eight frequency chunks) that are, by construction, zero for the reference antenna. 
These phase offsets must later be solved for using the polarization calibrator, but without applying the bandpass phases to the data.

A modification to this original scheme concerns the treatment of the bandpass phase solutions, which are no longer forced to zero. 
While the APS is active, the online phasing solution for each of the eight channel groups is referenced to the central frequency of that group. 
Because this central solution is not necessarily perfect for the edge channels, a phase slope appears across each group, causing decorrelation at the spectral edges. 
As a result, the amplitude bandpass solutions exhibit characteristic “eight-hump” structures, deviating from those of non-APS scans (e.g., see Fig. 10 in \citealt{Goddi2019} and Fig. 9 in \citealt{Crew2023}). 
Correcting this effect significantly improves the overall calibration quality, although the magnitude of improvement depends on the phasing efficiency, which in turn is influenced by weather, array configuration, and source properties.

In the current implementation, the APS bandpass solutions are computed toward the polarization calibrator. 
Ideally, an independent APS bandpass solution would be obtained for each actively phased target—since they are observed under different conditions of elevation, opacity, and atmospheric stability. 
However, because some targets display strong spectral absorption features (see Sections ~\ref{flag_sgra}, \ref{subsection:torus} and~\ref{subsection:ismSgra}), we restricted the solution to the polarization calibrator to ensure spectral integrity.
}

\subsubsection{Polarization corrections during APS reference antenna changes}

{The APS applies phase corrections to the individual antenna signals relative to a designated reference antenna~\citep{Matthews2018}. 
In rare cases, the reference antenna changes during an observing track. 
During the QA2 calibration, residual XY phase offsets and delays are corrected in the polarization calibration step (see Sect.\ref{polcal}) at the specified reference antenna; however, a change of this antenna introduces a substantial residual XY phase offset in the calibrated ALMA visibilities, which propagates into the VLBI data stream. 
To mitigate this effect, we implemented an additional correction routine that automatically detects any APS reference antenna change (or adopts a predefined reference if specified), computes the corresponding XY cross-polarization phase adjustments, and records them in a time-dependent polarization bandpass table. 
This procedure ensures a consistent instrumental polarization solution across the entire observing track, even when the APS reference antenna is switched mid-observation.}

\subsubsection{Flagging \sgra\ absorption lines}
\label{flag_sgra}
During the 2017 VLBI campaign, \citet{Goddi2021} reported the detection of absorption features in the  correlation spectra of \sgra, centered around $226.91\ \textnormal{GHz}$.
 These absorption features, first reported in Band 6,   are now consistently seen in Bands 3 and 7. Notable lines include HCN (88.6 GHz), HCO$^+$ (89.2 GHz), CS (98.0 GHz), and CN (226.6 GHz). Similar absorption is seen toward the nearby  QSO J1744--3116.

These absorptions cause flux suppression of a few percent, depending on the spectral window. To ensure consistency across \internaleht{SPW} in continuum polarization analysis, we flagged affected frequency channels prior to  imaging and  polarization analysis.

\subsubsection{Calibration script versioning and data releases}
\label{ScriptVersioning}
{Our calibration adopts the QA2 calibration script suite developed for ALMA Cycle~8, using \texttt{CASA~6.6}. This supersedes the earlier reductions in \citet{Goddi2019, Goddi2021}, which were based on \texttt{CASA~4.7} and Cycle~4 tools. While we generally follow the QA2 guidelines, we introduce several modifications---such as using alternative flux, bandpass, and polarization calibrators---to ensure cross-track consistency within each observing campaign.  Additional data flagging is also performed, based on both visual inspection of visibility amplitudes and phases and automated flagging routines (see Sect.\ref{subsection:flagging}). As a result, our calibrated datasets may differ slightly from the standard QA2 products archived at the ALMA Science Archive. For instance, we reprocessed the 2017 data originally analyzed in \citet{Goddi2021}, applying revised $T_{\mathrm{sys}}$ corrections, updated calibrator selections, and refined data flagging.  
These changes introduce minor variations in the derived Stokes fluxes and polarization fractions, all within the 1$\sigma$ (statistical and systematics) uncertainties. However, for quantities derived from fitting Stokes parameters vs frequency---such as Faraday rotation measures or depolarization ratios (see Sect.\ref{subsection:pol_imaging})---where the uncertainties are obtained from least-squares fits across the four SPWs, results for some sources may differ by more than 1$\sigma$. In such cases, the newly derived values supersede those reported in \citet{Goddi2021}, which were based on the earlier calibration  \citep{Goddi2019}.  
Finally, as new releases of the QA2 script suite incorporate improved calibration procedures or new functionalities (e.g. refined flux calibration; see Sect.~\ref{app:fluxscale}), we plan to reprocess the affected datasets and issue updated VAPOLA data releases superseding previous ones (see Appendix~\ref{data_releases}).  
}

\subsection{Pipeline for data analysis and repository}\label{sec:pipeline}

The calibration process described in Sect.~\ref{sec:qa2} yields two types of Measurement Set ~\citep[MS][]{Kemball2000} files including all sources observed in a VLBI track: a full-resolution MS for frequency-domain and spectral line studies, and a frequency-averaged MS optimized for continuum and polarization analysis.

\begin{figure*}[ht]
\centering
\resizebox{\textwidth}{!}{%
\begin{tikzpicture}[
    font=\sffamily,
    node distance=4em,
    line/.style={-{Latex}, thick},
    dashedline/.style={-{Latex}, dashed, thick},
    data/.style={
      cylinder,
      cylinder uses custom fill,
      shape border rotate=90,
      aspect=0.5,
      draw,
      minimum height=3em,
      text width=5em,
      text centered
    },
    block/.style={
      rectangle,
      draw,
      rounded corners,
      align=center,
      text width=7em,
      minimum height=3em
    },
    product/.style={
      rectangle,
      draw,
      align=center,
      text width=9em,
      rounded corners,
      minimum height=3em
    },
    cloud/.style={
      draw,
      ellipse,
      align=center,
      minimum height=2em,
      text width=4em
    }
]

\node[data](qa2){QA2\\DATA};
\node[block, right=8em of qa2](flag){FLAGGING};
\node[block, right=8em of flag](imag){IMAGING};
\node[block, right=8em of imag](pol){POLARIMETRY};
\node[cloud, right=8em of pol](sgra){Sgr A*};

\node[product, above=4em of imag](stokes){
  \textbf{Products}\\
  Full Stokes Images\\
  (per and combined \internaleht{SPW})
};
\node[product, above=4em of pol, yshift=+1.2em](polmaps){
  \textbf{Products}\\
  Polarization maps\\
  Polarimetric\\tabulated products
};
\node[product, above=4em of sgra](sgraProd){
  \textbf{Products}\\
  Light curves\\
  Periodograms\\
  Transients
};

\draw[line] (qa2) -- (flag);
\draw[line] (flag) -- (imag);
\draw[line] (imag) -- (pol);
\draw[line] (pol) -- node[above]{If} (sgra);

\path (qa2) -- (flag) coordinate[pos=0.3](mid);
\node[block, above=5em of mid](conf){config};
\draw[line] (conf.south) |- (mid);

\draw[dashedline] (imag.north) -- (stokes.south);

\draw[dashedline] (pol.north) -- (polmaps.south);

\draw[line] (sgra) -- (sgraProd);

\begin{pgfonlayer}{background}
\node[
  draw,
  shape=ellipse,
  fit=(stokes)(polmaps)(sgraProd),
  inner sep=.5em,
  label=above:{VAPOLA}
] {};
\end{pgfonlayer}

\end{tikzpicture}
}
\caption{Schematic overview of the VAPOLA processing pipeline. Starting from the QA2-calibrated ALMA data (left), the pipeline applies additional automated flagging and full-Stokes imaging steps before producing science-ready products. These include (i) full-Stokes images for individual and combined (0+1),(2+3) spectral windows, (ii) polarimetric products such as polarization maps and visibility-domain Stokes fits for compact sources, and (iii) for tracks containing \sgra, dedicated time-domain products.}
\label{fig:pipeline}
\end{figure*}
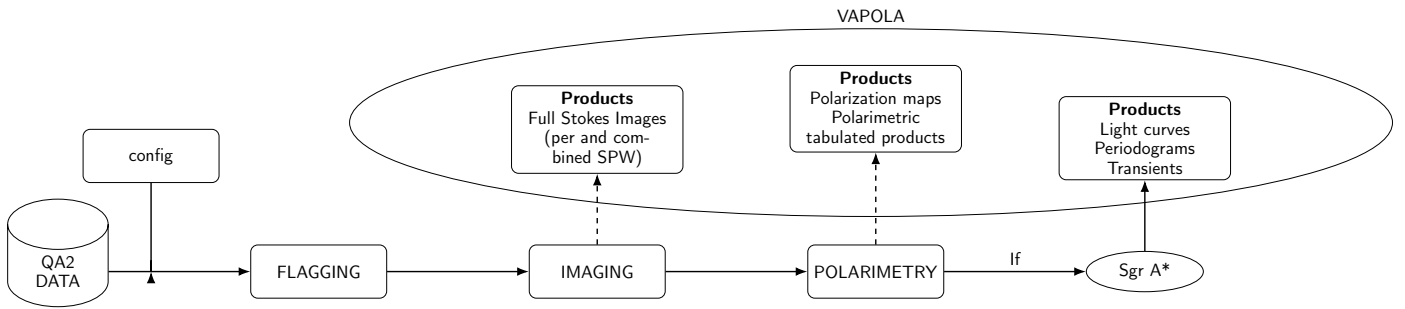

We have developed an automated pipeline to process the fully calibrated MS data, generating science-ready, high-level data products. Figure~\ref{fig:pipeline} illustrates the schematic data flow and processing stages.

For each track, the pipeline performs the following operations:
\begin{itemize}
  \item Generation of diagnostic plots and metadata summaries (Sect.~\ref{subsection:diag}).
  \item Final flagging of visibility outliers not removed during QA2 (Sect.~\ref{subsection:flagging}).
       \item Modeling of compact sources in the $uv$ domain using \texttt{UVMULTIFIT}  
\citep{uvmultifit}, and derivation of their polarimetric quantities (Sect.~\ref{subsec:uvmf}).
  \item Full-Stokes imaging for each SPW, and combined imaging across SPWs (Sect.~\ref{subsection:imaging}).
   \item Combination of Stokes parameters to construct  spectral and polarization maps 
   \ref{subsection:pol_imaging}).
 \end{itemize}

If \sgra\ is present in a track, we apply a specialized calibration strategy to address its intrinsic variability. This includes the minispiral calibration technique \citep{Wielgus2022, mus22, Albentosa2025}, followed by time-domain analysis and generation of full-Stokes light curves per \internaleht{SPW} (Sect.~\ref{subsubsec:sgra}).

\subsubsection{Diagnostics \& summarizing the data}
\label{subsection:diag}

The pipeline operates on a per-track basis, where each ``track'' is a \texttt{CASA} MS file from a single observing session. While retrieving basic metadata such as date, band, and source list can be done manually, it becomes cumbersome when scaling to large datasets. To enable efficient inspection, the pipeline generates a summary text file (\texttt{summ}) for each track.

Each \texttt{summ} file includes:
\begin{itemize}
  \item Track name and corresponding MS filename;
  \item Observing mode (VLBI or non-VLBI);
  \item ALMA band, reference frequency, and wavelength;
  \item Baseline length in meters, (minimum distance / maximum distance).
  \item Reference date, and starting and ending days;
  \item For each science target: identifier, source name, net integration time (\texttt{Tint}), and number of antennas involved.
\end{itemize}

The integration time \texttt{Tint} reflects the calibrated visibilities after flagging  and may differ from the total on-source time during observations. Additionally, a standardized source naming convention is enforced across the data repository in order to ease data discovery, and to ensure consistency over multiple years and bands. 

This information is shared in a form of a table (more details will be given in Sect.~\ref{sec:architecture}) hosted on the VAPOLA webpage. 

\subsubsection{Flagging}
\label{subsection:flagging}

Despite QA2 calibration, residual bad data may remain—typically due to  interference, electronic glitches, or sub-optimal phasing solutions (typically in the first integration of each VLBI scan; \citealt{Goddi2019}).

To minimize manual inspection, the pipeline incorporates automated flagging tools. Two strategies are available:

\paragraph{CUTOFF method:} All data with visibility amplitudes or phases outside the range $\bar{\mathrm{Amp}} (\bar{\mathrm{Pha}} ) \pm k\sigma$ are flagged, \internaleht{where the overlined $\mathrm{Amp}$ and $\mathrm{pha}$ denote the median amplitude  and phase}, $\sigma$ the standard deviation, and $k$ a configurable threshold (typically 3 or 4).

\paragraph{Isolation Forest:} This machine learning algorithm learns the distribution of complex visibilities and identifies outliers as points that are relatively isolated in the adopted feature space \citep{IsolationForest}. Unlike the CUTOFF method, it operates on the full complex plane (real and imaginary components jointly), allowing for the identification of anomalous visibilities that may not be captured by simple amplitude or phase thresholds (see Appendix~\ref{app:isolation_forest} for more details). 
By default, the pipeline uses the Isolation Forest method with a contamination factor of up to 5\%, which should be understood as an upper limit on the additional fraction of data that may be flagged by this procedure, rather than the expected fraction of valid data removed.

We emphasize that the flagging step method is applied conservatively. In addition, it is only applied to sources that are compact and not dominated by extended structure; for sources such as \mes and \sgra, no additional automated flagging is done beyond QA2 at this stage.

\subsubsection{Polarimetric tables}\label{subsec:uvmf}

The VAPOLA repository also includes tabulated Stokes parameters and derived polarimetric quantities for the compact cores of all sources. These are extracted from the visibility data using the \texttt{UVMULTIFIT} package\footnote{\url{https://github.com/marti-vidal-i/UVMultiFit}} \citep{uvmultifit}, an external \casa\ tool designed for model-fitting in the $uv$ domain.

\texttt{UVMULTIFIT} performs non-linear least-squares minimization to fit model visibilities to the observed ones, solving:
\begin{equation}
    \chi^2 = \sum_{i,j} \frac{|V^{\mathrm{mod}}_{i,j} - V^{\mathrm{obs}}_{i,j}|^2}{\sigma_{i,j}},
\end{equation}
where $V^{\mathrm{mod}}$ and $V^{\mathrm{obs}}$ are the model and observed visibilities for baseline $(i,j)$, and $\sigma_{i,j}$ is the associated uncertainty.

While \texttt{UVMULTIFIT} can accommodate a variety of geometric models (e.g., rings, Gaussians), we assume that the mm/sub-mm source emission is dominated by a compact component at the phase center. This assumption is motivated by the fact that most AGN targets are compact on arcsecond scales and/or are dominated by a bright unresolved core. Moreover, \citet{Goddi2021} (Appendix~C) quantified the effect of extended emission in sources that do show structure on arcsecond scales (notably M87 and \sgra) on flux estimates derived in the visibility domain. They found that deviations in the recovered Stokes IQUV values, when comparing uv-domain fitting to image-based methods that explicitly account for source structure, remain well within $1\sigma$. Based on this, we model the data with a delta function (point source) fitted independently to the visibilities of each spectral window, and we report the corresponding best-fit Stokes parameters. 
Polarimetric quantities such as LP, CP and zero-wavelength EVPA are computed from the fitted Stokes values, and tabulated per source, band, and epoch. 
The quoted uncertainties include a systematic uncertainty of 0.03\% of Stokes~$I$ in Stokes~$Q$ and $U$, corresponding to one third of the ALMA minimum detectable linear polarization for compact sources (0.1\% of Stokes~$I$ at the $3\sigma$ level; see the ALMA Technical Handbook, \citealt{ALMAHandbook13}). For Stokes~$V$, we adopt a systematic uncertainty of 0.6\% of Stokes~$I$, corresponding to one third of the ALMA minimum detectable circular polarization for compact sources (1.8\% of Stokes~$I$ at the $3\sigma$ level). The total uncertainties are computed by combining the thermal and systematic contributions in quadrature.
All tabulated results from \texttt{UVMULTIFIT} are included in the final data products (Sect.~\ref{subsubsec:tables}).

\subsubsection{Imaging}\label{subsection:imaging}

The next step in the pipeline is imaging, performed automatically using the CLEAN deconvolution algorithm \citep{Hogbom1974,Clark1980} via the \texttt{tclean} task in \casa.

For each source in every track, we produce full-Stokes images for: all four individual spectral windows (SPWs 0, 1, 2, 3), the two sidebands (combining SPWs 0+1 and 2+3), and the full-band combination (all four SPWs). For the individual SPWs and sidebands, we use \texttt{deconvolver='hogbom'} with \texttt{nterms=1}. For the combined SPW image, we adopt the multi-term multi-frequency synthesis algorithm (\texttt{deconvolver='mtmfs'}) with \texttt{nterms=2} \citep{Rau2011}. This results in seven images per source per epoch, later used for polarimetric analysis.

Images are generated with Briggs weighting \citep{Briggs1995} using a robust parameter of 0.5 and a cleaning gain of 0.1. Pixel sizes and image dimensions vary depending on the band and on the array configuration during each observation and are selected to fully cover the ALMA primary beam: $\sim$60\arcsec\ for Band 3, $\sim$27\arcsec\ for Band 6, and $\sim$18\arcsec\ for Band 7.

To automate the CLEANing process, we implement a dynamic masking script for each Stokes parameter. Then, for every Stokes, same mask is used for the four SPWs. The procedure begins with an initial shallow CLEAN (100 iterations) restricted to an inner region—typically 10\arcsec\ (Band 3) or 4\arcsec\ (Band 6). Beam parameters for the individual and sideband images are standardized to match those of the full-band image using the \texttt{restoringbeam} keyword.
 If residual emission exceeding $7\sigma$ is detected in the residual maps from the first round of cleaning, the script adjusts the clean mask accordingly, and a second  round of cleaning is executed down to a $2\sigma$ threshold. Both cleaning stages are performed with \texttt{interactive=False}, although in selected cases with complex structure (e.g., \mes), an additional \texttt{interactive=True} step is manually invoked. 
 No additional self-calibration is performed at this stage. The QA2 calibration of ALMA VLBI data already includes antenna-based gain solutions derived from the observed targets, effectively incorporating a form of self-calibration during earlier processing steps \citep{Goddi2019}. We verified that applying standard post-imaging self-calibration does not lead to significant improvements in image quality or dynamic range, even for sources with extended emission such as M87. An exception is made for \sgra, whose intrinsic variability requires dedicated self-calibration (see Sect..~\ref{subsubsec:sgra}).

The resulting full-Stokes images are stored in both \casa\ and FITS formats.

\subsubsection{Spectropolarimetric imaging}\label{subsection:pol_imaging}

The full-Stokes images from each of the four SPWs are used to construct spatially resolved maps of spectral and polarization properties for extended sources. These include:

\begin{itemize}
    \item \textbf{Spectral Index} ($\alpha$ or ALPHA): derived from the first-order Taylor term of the full-band \texttt{mtmfs} image.
    \item \textbf{Polarized Intensity} ($I_p$): linear polarization flux in Janskys, computed as $\sqrt{Q^2+U^2}$ from the full-band image.
    \item \textbf{Linear Polarization} (LP) fraction: computed as $\sqrt{Q^2+U^2}/I$ for the full-band image.
    \item \textbf{Circular Polarization} (CP) fraction: computed as $V/I$ for the full-band image.
    \item \textbf{Depolarization} (Dep): in units of $10^{-4}$ GHz$^{-1}$, obtained via a pixel-wise linear fit of LP as a function of frequency ($\nu$) across the four SPWs.
    \item \textbf{Electric Vector Position Angle} (EVPA or $\chi$): in degrees, calculated as $\chi = 0.5\,\arctan(U/Q)$ for the full-band image.
    \item \textbf{Rotation Measure} (RM): in units of $10^{5}$ rad/m$^{2}$, obtained from a pixel-wise linear fit of $\chi$ versus $\lambda^2$ of each individual \internaleht{SPW}.
    \item \textbf{EVPA at Zero Wavelength} ($\chi_0$): in degrees, the y-intercept of the $\chi$ vs.\ $\lambda^2$ fit.
\end{itemize}

Error maps are also produced for each of these quantities. For derived ratios (LP, CP, EVPA), we propagate uncertainties by combining thermal noise with instrumental leakage terms in quadrature. As reported in \citet{Goddi2021}, leakage contributes $\sim$0.03\% of $I$ to $Q$ and $U$, and $\sim$0.6\% of $I$ to $V$. For quantities derived from fitting (e.g., $\alpha$, RM, Depol., $\chi_0$), the errors are computed from the least-squares fit over the four different SPWs.

We stress that the systematic uncertainties quoted in Sect.~\ref{subsec:uvmf} refer only to measurements at the phase center. ALMA linear polarization performance is formally guaranteed at the 0.1\% level of Stokes~$I$ only within the inner one-third of the primary beam. For emission extending beyond this region, systematic errors increase as a result of instrumental polarization effects and may reach $\sim$0.3--0.5\% of Stokes~$I$ near the primary-beam FWHM \citep{Hull2020}. Therefore, users should exercise caution when interpreting low-level polarization signals outside the inner one-third of the primary beam and should consider these larger systematic uncertainties when assessing the significance of polarized emission. To facilitate this assessment, the pre-plotted maps described in Sect.~\ref{plotted_maps} include a dashed circle marking the boundary of the inner one-third of the primary beam.

\subsubsection{Dedicated procedures for \sgra}\label{subsubsec:sgra}

Unlike most VLBI targets, \sgra\ exhibits intrinsic variability on short timescales (minutes), as shown in recent time-domain studies \citep[e.g.][]{macquart2006, marrone2008, dexter2014, Wielgus2022, mus22, vonFellenberg2023}. This variability presents challenges for standard interferometric calibration, but also provides a unique opportunity to probe the dynamics of plasma near a supermassive black hole.

To accommodate this behavior and enable time-resolved analysis, the pipeline includes two dedicated components:
\begin{itemize}
    \item A specialized calibration method, known as \textit{minispiral calibration}, designed to handle the source variability \citep{Wielgus2022, mus22, EHTmemo}.
    \item A routine to extract time-domain light curves for the mm emission in full Stokes parameters.
\end{itemize}

Because time coverage is crucial for these studies, the automatic flagging procedure described in Sect.~\ref{subsection:flagging} is disabled for \sgra. Only visibilities with antenna elevations below $25^\circ$ are flagged. This maximizes $uv$-coverage and ensures the best possible temporal fidelity in the derived light curves.

Below, we outline the procedure followed to obtain the \sgra\ light curve products  \citep[see][for details]{EHTmemo}.

\vspace{0.5em}
\paragraph{Minispiral calibration.} 
This technique decomposes the observed emission into two components:
\begin{enumerate}
    \item a bright, compact, intrinsically time-variable component associated with \sgra;
    \item an extended, quasi-stationary component (the ``minispiral'') that acts as a local reference structure.
\end{enumerate}

By modeling and subtracting the extended emission, we isolate the compact, variable signal of \sgra. A time-dependent complex gain is then applied at each integration to account for its intrinsic variability. This gain is determined by minimizing the chi-squared statistic:
\begin{equation}\label{eq:fitting}
\chi_t^2\left(F_{\text{Sgr A}^*}, F_{j, t}^{\text{ext}}\right),
\end{equation}
where $F_{\text{Sgr A}^*}$ is the flux density of the compact source, and $F_{j, t}^{\text{ext}}$ represents the modeled extended emission on  baseline $j$ at time $t$. 

We note that the standard self-calibration assumption of a constant-flux
central source would instead enforce a fixed core brightness
($F_{\text{Sgr A}^*}$), artificially transferring all intrinsic variability
to the minispiral component ($F_{j,t}^{\text{ext}}$). By applying the
derived time-dependent gain, this spurious effect is removed and the
variability is correctly attributed to the compact source.

\vspace{0.5em}
\paragraph{Workflow overview.}
The \sgra\ processing pipeline proceeds as follows:
\begin{enumerate}
    \item A first CLEAN is performed on the full emission (core + minispiral) to build an initial model.
    \item CLEAN components corresponding to \sgra\ are subtracted, leaving a model of the extended emission.
    \item A model-fitting step is then applied to the visibilities using the residual model, solving for the compact flux at each time step using the minispiral calibration method.
    \item Time-domain light curves are extracted for each spectral window and for each Stokes parameter (I, Q, U, V).
    \item A final outlier removal step is applied using the Isolation Forest algorithm, cleaning the light curves from residual calibration artifacts \citep{EHTmemo}.
\end{enumerate}

\vspace{0.5em}
\paragraph{Output products.}
For each spectral window, the final data products generated include:
\begin{itemize}
    \item time-resolved light curves of Stokes I, LP,  and EVPA;
    \item CLEAN component lists and CLEAN masks;
    \item flag version tables used in the calibration.
\end{itemize}

These products are essential for the analysis of the variable polarization and intensity structure of \sgra, and are made available in the VAPOLA data collection.

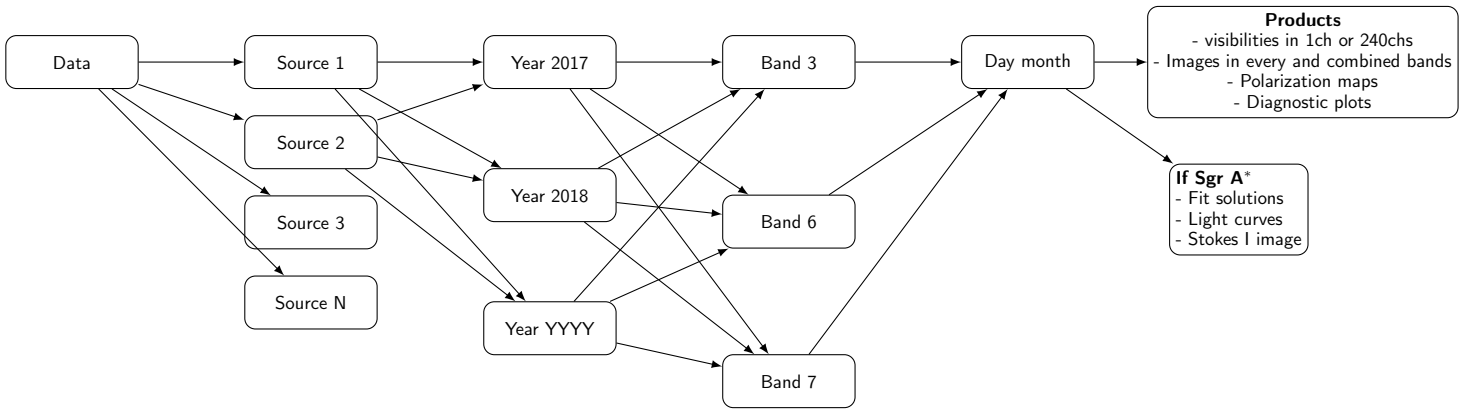
\begin{figure*}
\begin{tikzpicture}[
  scale=0.7,          
  transform shape,     
  font=\sffamily,
  >=latex,
  node distance=1.2cm and 2.0cm,
  box/.style={
    rectangle,
    draw=black,
    rounded corners,
    align=center,
    minimum width=2.5cm,
    minimum height=1cm
  },
  data/.style={box, fill=white},
  source/.style={box, fill=white},
  year/.style={box, fill=white},
  band/.style={box, fill=white},
  daymonth/.style={box, fill=white},
  products/.style={box, fill=white}
]

\node[data] (repository) {Data};

\node[source] (s1) [right=2.0cm of repository] {Source 1};
\node[source] (s2) [below=0.5cm of s1]  {Source 2};
\node[source] (s3) [below=0.5cm of s2]  {Source 3};
\node[source] (sn) [below=0.5cm of s3]  {Source N};

\node[year] (y2017) [right=2.0cm of s1] {Year 2017};
\node[year] (y2018) [below=1.5cm of y2017] {Year 2018};
\node[year] (y2019) [below=1.5cm of y2018] {Year YYYY};

\node[band] (band1) [right=2.0cm of y2017] {Band 3};
\node[band] (band2) [below=2.0cm of band1] {Band 6};
\node[band] (band3) [below=2.0cm of band2] {Band 7};

\node[daymonth] (daymonth) [right=2.0cm of band1] {Day month};

\node[products] (prod) [
  right=0.8cm and 1.0cm of daymonth,
] {
  \textbf{Products}\\
  - visibilities in 1ch or 240chs\\
  - Images in every and combined bands\\
  - Polarization maps\\
  - Diagnostic plots
};

\node[products] (ifsgra) [
  below right=2cm of daymonth,
  align=left
] {
  \textbf{If \sgra}\\
  - Fit solutions\\
  - Light curves \\
  - Stokes I image
};

\draw[->] (repository) -- (s1);
\draw[->] (repository) -- (s2);
\draw[->] (repository) -- (s3);
\draw[->] (repository) -- (sn);

\draw[->] (s1) -- (y2017);
\draw[->] (s1) -- (y2018);
\draw[->] (s1) -- (y2019);

\draw[->] (s2) -- (y2017);
\draw[->] (s2) -- (y2018);
\draw[->] (s2) -- (y2019);

\draw[->] (y2017) -- (band1);
\draw[->] (y2018) -- (band1);
\draw[->] (y2019) -- (band1);

\draw[->] (y2017) -- (band2);
\draw[->] (y2018) -- (band2);
\draw[->] (y2019) -- (band2);

\draw[->] (y2017) -- (band3);
\draw[->] (y2018) -- (band3);
\draw[->] (y2019) -- (band3);

\draw[->] (band1) -- (daymonth);
\draw[->] (band2) -- (daymonth);
\draw[->] (band3) -- (daymonth);

\draw[->] (daymonth) -- (prod);
\draw[->] (daymonth) -- (ifsgra);

\end{tikzpicture}
\caption{Logical structure of the VAPOLA archive. For each source, data products are organized hierarchically by observing year, ALMA band, and observing date. At the lowest level, every date directory contains fully calibrated visibilities (single-channel and full-frequency), full-Stokes images, polarization maps, and diagnostic plots. For \sgra, additional subdirectories provide specialized time-domain products derived from the minispiral-calibrated data.}
\label{fig:archive_architecture}
\end{figure*}

\section{VAPOLA's architecture}\label{sec:architecture}

The repository follows a robust and well-defined architecture designed to ensure systematic data access and maximize usability. At the top level, the root directory (\texttt{./}) contains two main subdirectories, \texttt{Data} and \texttt{Tables}, which organize the products according to their type. In the following, we describe the structure and content of these directories.

\subsection{Tables}\label{subsubsec:tables}

This directory provides structured and summarized information about the observations and the derived data products. Its primary objective within VAPOLA is to give users quick and user-friendly access to key information.

The \texttt{Tables} folder is divided into two subdirectories:  
\begin{itemize}
    \item \texttt{observations\_summary}: contains per–track summaries. For each frequency band, the tables include the track name, observation date, number of phased antennas,  the minimum/maximum projected baseline length (in meters), the observing start and end times. Each track entry lists one or more ALMA project code(s) observed within that track, the observed source (with the possibility of multiple entries per source when different projects are involved), the observing mode (V: VLBI; NV: non-VLBI), and the net cumulative on-source integration time (after calibration and flagging). This corresponds to Table~\ref{tab:list_projects}.  
    \item \texttt{polarization\_tables}: contains yearly CSV tables of the polarization values described in Sect.~\ref{subsec:uvmf}, organized by observation date/track. Each table lists the Stokes parameters ($I, Q, U, V$) with their associated errors ($I_{\mathrm{e}}, Q_{\mathrm{e}}, U_{\mathrm{e}}, V_{\mathrm{e}}$), the linear and circular polarization fractions (LP, CP) with uncertainties (LP$_{\mathrm{e}}$, CP$_{\mathrm{e}}$), and the EVPA with errors ($\chi$, $\chi_{\mathrm{e}}$). Values are reported for each of the four SPWs in every band (see Table E1 in~\citealt{Goddi2021} for an example).  
\end{itemize}

\subsection{Data}

Figure~\ref{fig:archive_architecture} presents an overview of the \texttt{Data}  repository layout. \texttt{Data} products are divided into two primary categories: VLBI and non-VLBI scans, each following the hierarchical structure illustrated in Fig.~\ref{fig:archive_architecture}.

For both VLBI and non-VLBI scan data, the directory tree is organized by observed source, then further subdivided by year, observing band, and observing date (day and month). At the date level, a standard set of subdirectories is created:

\begin{itemize}
\item \texttt{diagnostics}
\item \texttt{images}
\item \texttt{plots}
\item \texttt{pol\_products} 
\item \texttt{visibilities}
\item  \texttt{light_curves} (only for \sgra)
\end{itemize}

Each subdirectory contains a specific category of high-level data products. Below, we describe in detail the contents of each folder. We refer the reader to Appendix~\ref{app:vapola_download} for information on web access, user interaction and data versioning.

\subsubsection{Fully calibrated visibilities}

The visibilities directory provides fully calibrated visibility datasets in the standard CASA MS format. Those data are available in two forms:

\begin{itemize}
    \item \texttt{average\_frequency}: datasets averaged into a single spectral channel (filenames with the suffix \texttt{1ch}).  
    \item \texttt{full\_frequency}: datasets retaining all 240 channels, enabling detailed spectral analyses (see Sect.~\ref{sec:science_case}).  
\end{itemize}  

Each MS contains, in turn, a flagversion table (\texttt{.flagversion}) which store flagging information produced as described in Sect.~\ref{subsection:flagging}. These allow users to revert to previous flagging states if needed.

\subsubsection{Diagnostic plots}

The \texttt{diagnostics} folder contains sets of \texttt{.png} files showing visibility distributions and flagged data (highlighted in red), generated during the procedure described in Sect.~\ref{subsection:diag}. Each diagnostic plot consists of amplitude versus time and phase versus time panels, displayed for all four spectral windows and both XX and YY correlations.  
These plots allow users to inspect the quality of the observations and evaluate whether the applied flagging is appropriate. An example is shown in Fig.~\ref{fig:flag_diag} (see Appendix~\ref{app:isolation_forest} for details).

\subsubsection{Full-Stokes images}
\label{stokesim}

The \texttt{images} folder contains full-Stokes images for each SPW and for SPW-combined datasets.  
Images are provided in two formats: (1) the native \casa\ output format and (2) \texttt{FITS}.  

The \casa\ outputs include the standard \texttt{CLEAN} products: \texttt{image}, \texttt{mask}, \texttt{model}, \texttt{psf}, \texttt{residual}, and \texttt{sumwt}. Files with the extension \texttt{.image} (and associated products) must be read within \casa\ (see the \casa\ documentation for details).  
Alternatively, the corresponding \texttt{FITS} files can be downloaded directly from the \texttt{FITS} subfolder.

\subsubsection{Spectropolarimetric maps}

Polarization and spectral index maps are derived with \casa\ from the full-Stokes images produced in the imaging step as described in Sect.~\ref{subsection:pol_imaging} and stored in the \texttt{pol\_products} folder, together with their associated error maps and their residuals. Both \casa\ and \texttt{FITS} formats are available.  

To ensure that only robust polarization measurements are shown, signal-to-noise thresholds are applied:  
\begin{itemize}
    \item Spectral index: $I > 4\sigma$ in the combined SPW Stokes $I$ image.  
    \item LP, EVPA, depolarization, and RM: $I_p > 3\sigma$.  
    \item CP: $V > 5\sigma$.  
\end{itemize}  
We note that robust Stokes~$V$ detections are particularly challenging with these ALMA observations \citep[see discussion in Appendix G of][]{Goddi2021}.  

\subsubsection{Pre-plotted maps}
\label{plotted_maps}
\begin{figure*}
    \centering

    \includegraphics[width=0.31\linewidth, height=0.31\linewidth]{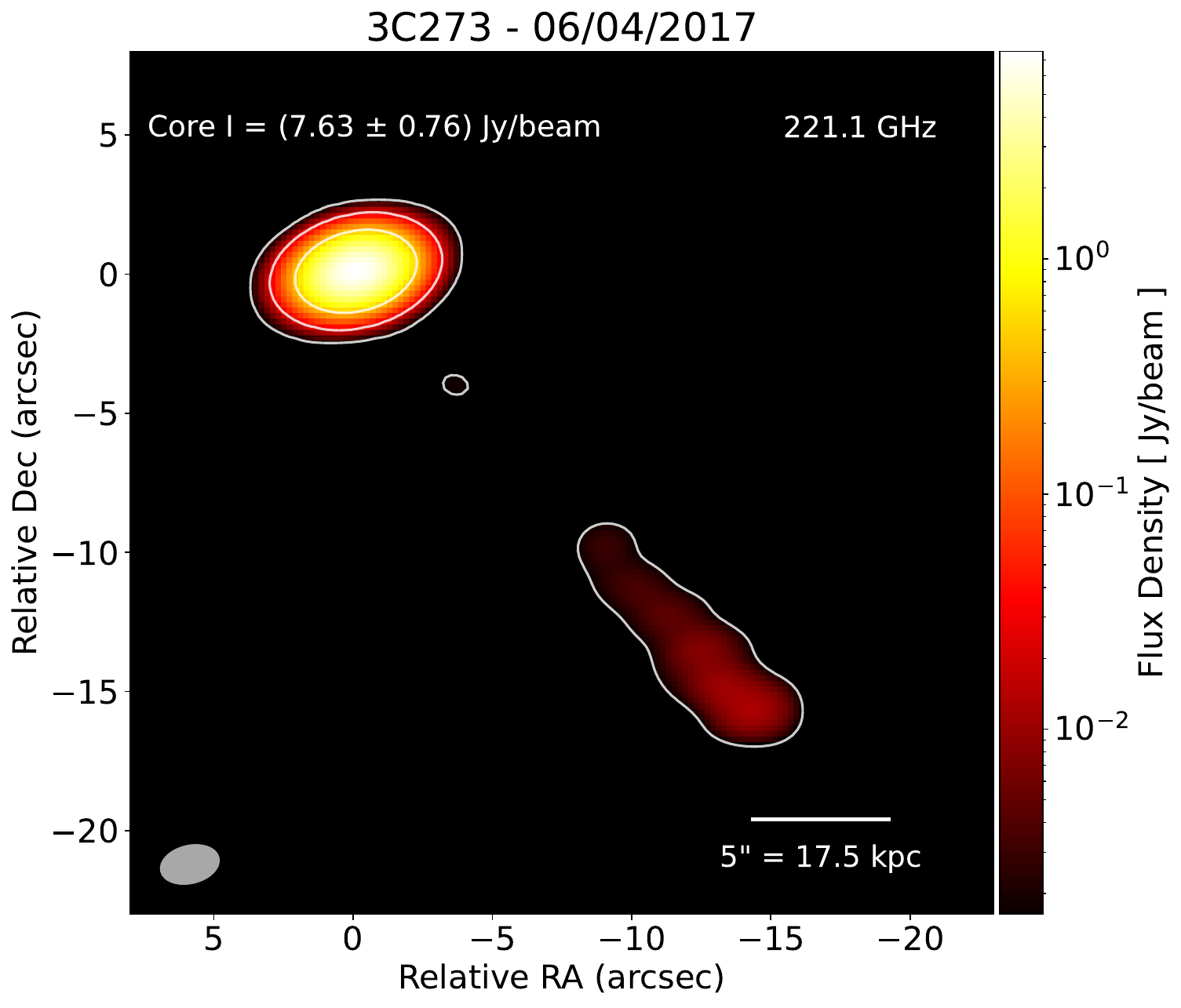}
    \includegraphics[width=0.31\linewidth, height=0.31\linewidth]{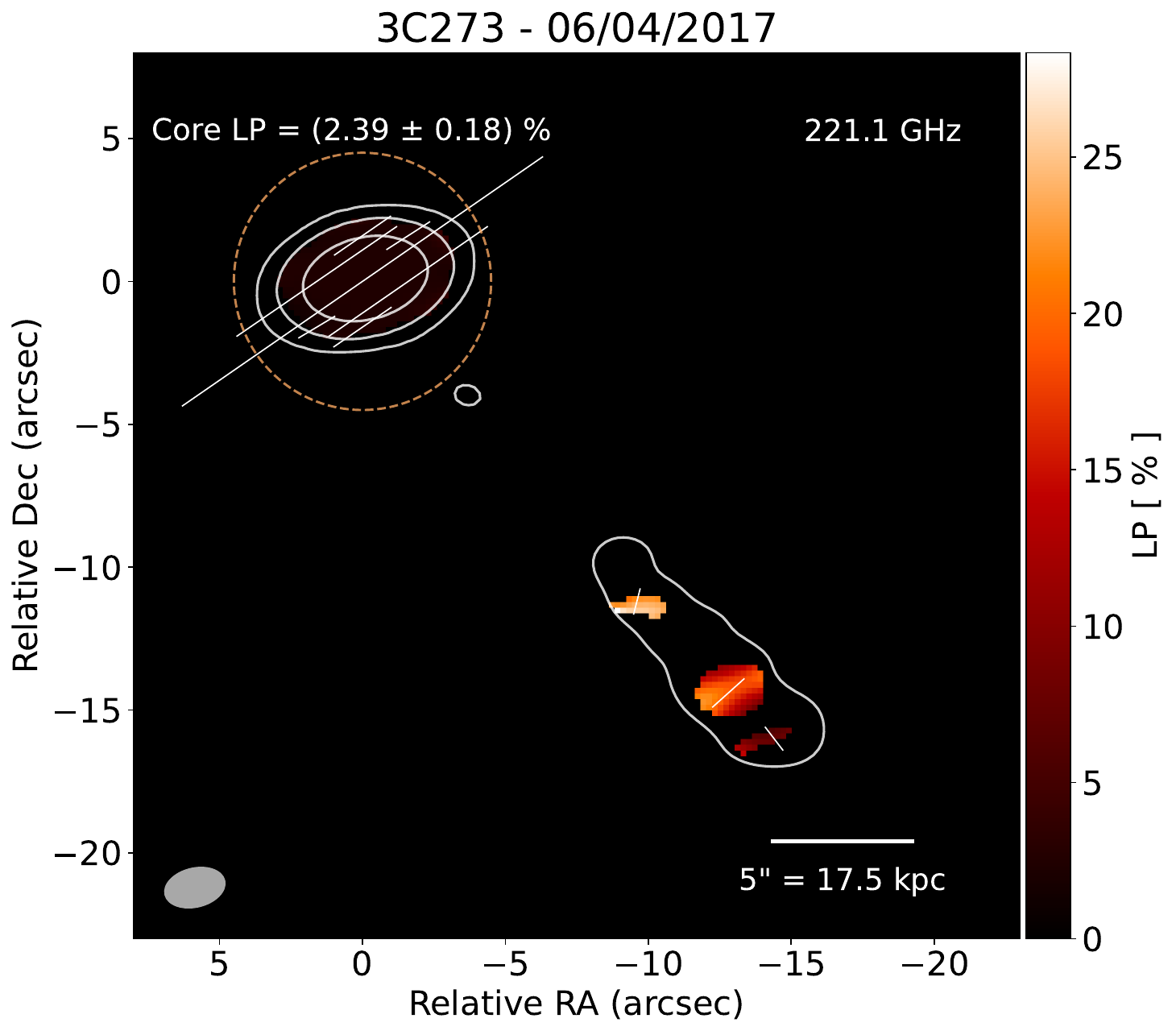}
    \includegraphics[width=0.31\linewidth, height=0.31\linewidth]{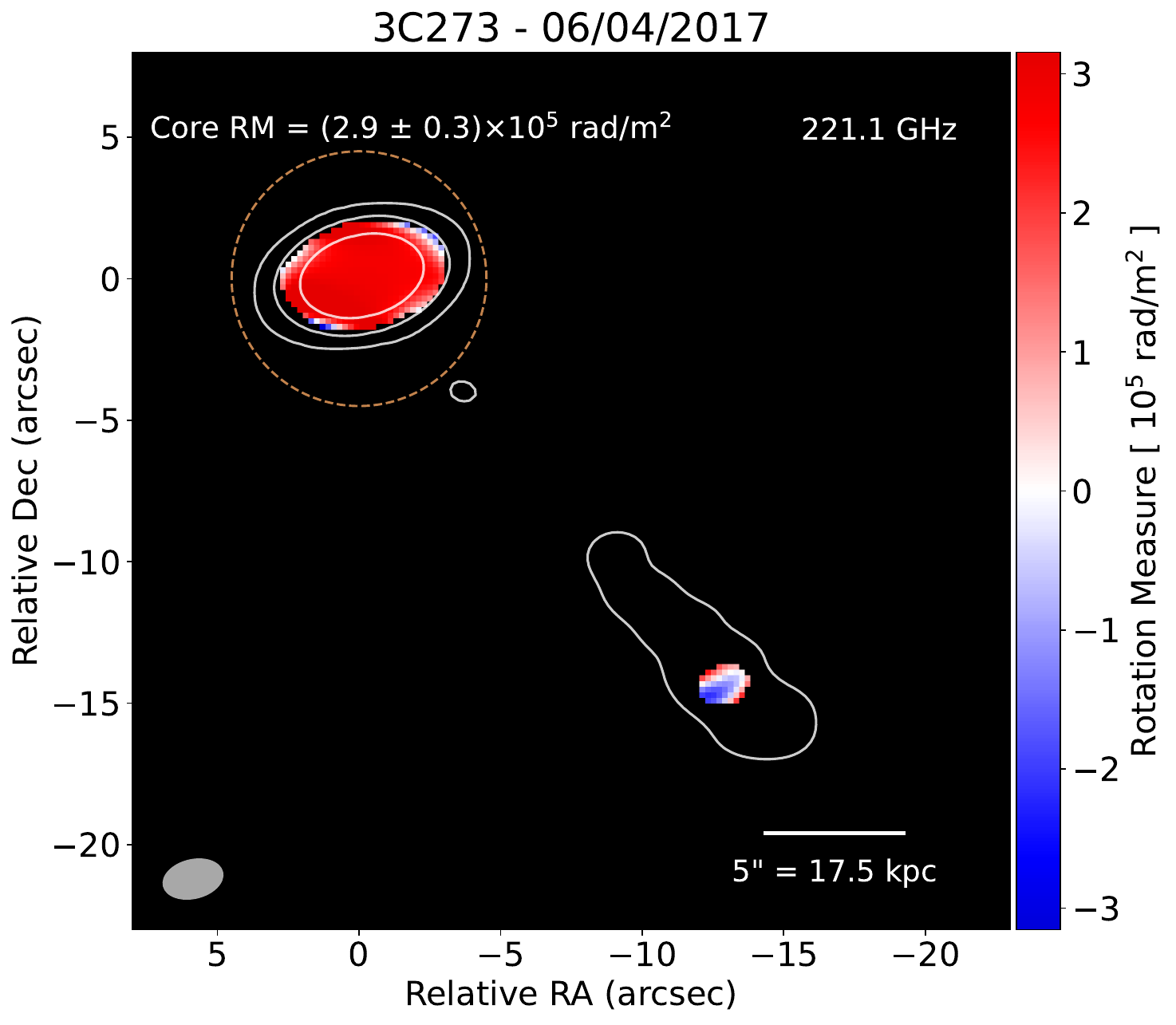} \\
    \includegraphics[width=0.31\linewidth, height=0.31\linewidth]{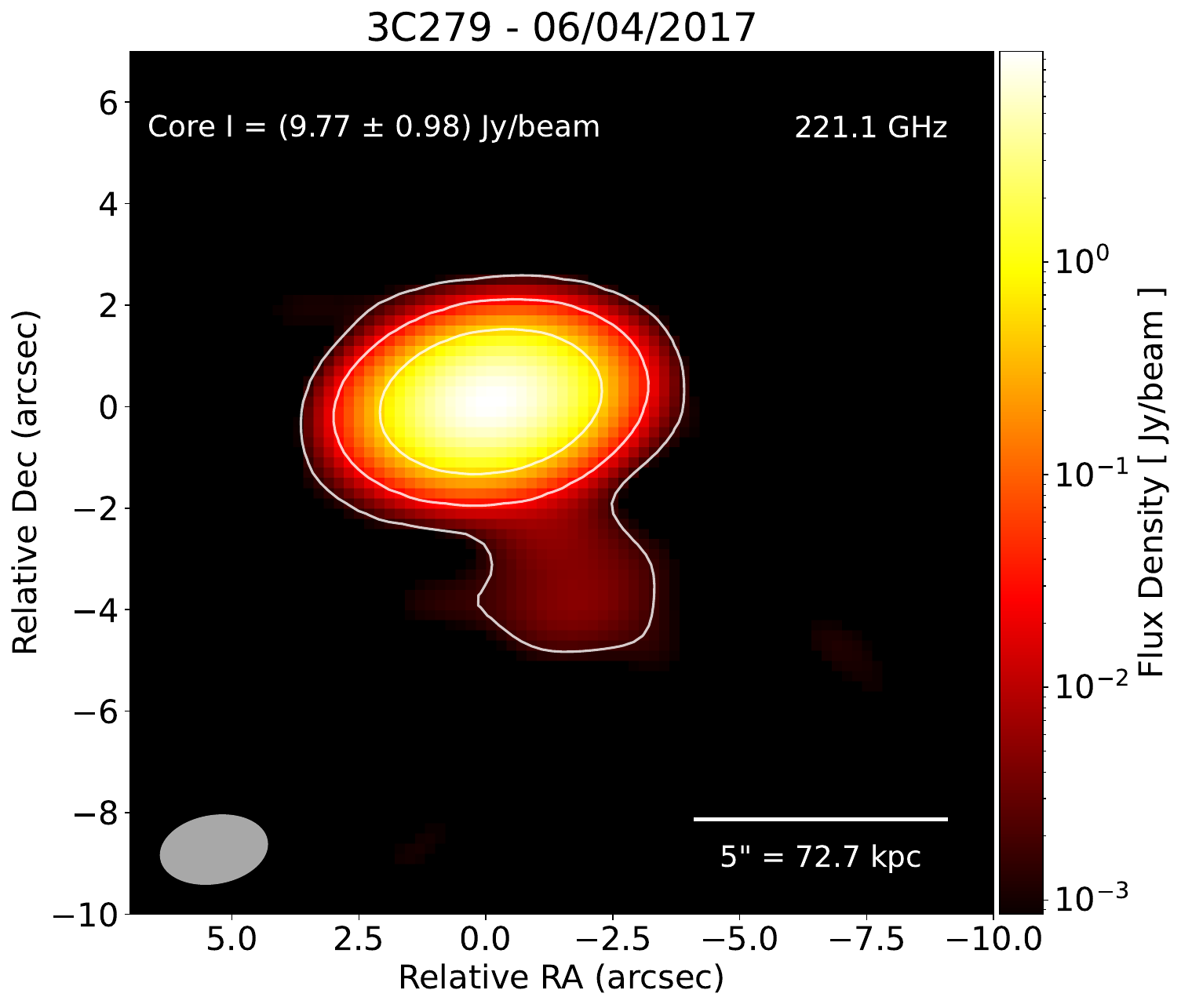}
    \includegraphics[width=0.31\linewidth, height=0.31\linewidth]{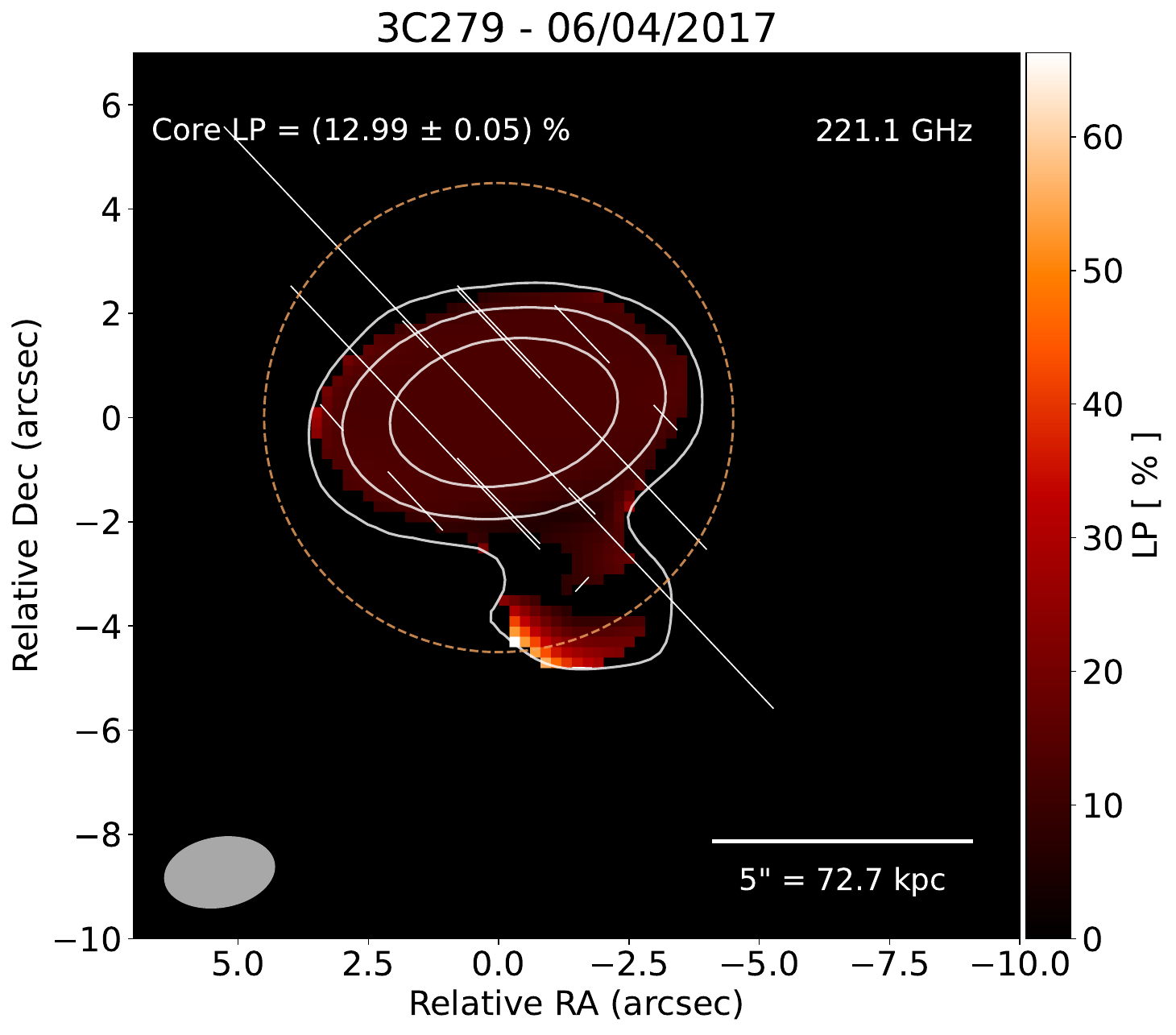}
    \includegraphics[width=0.31\linewidth, height=0.31\linewidth]{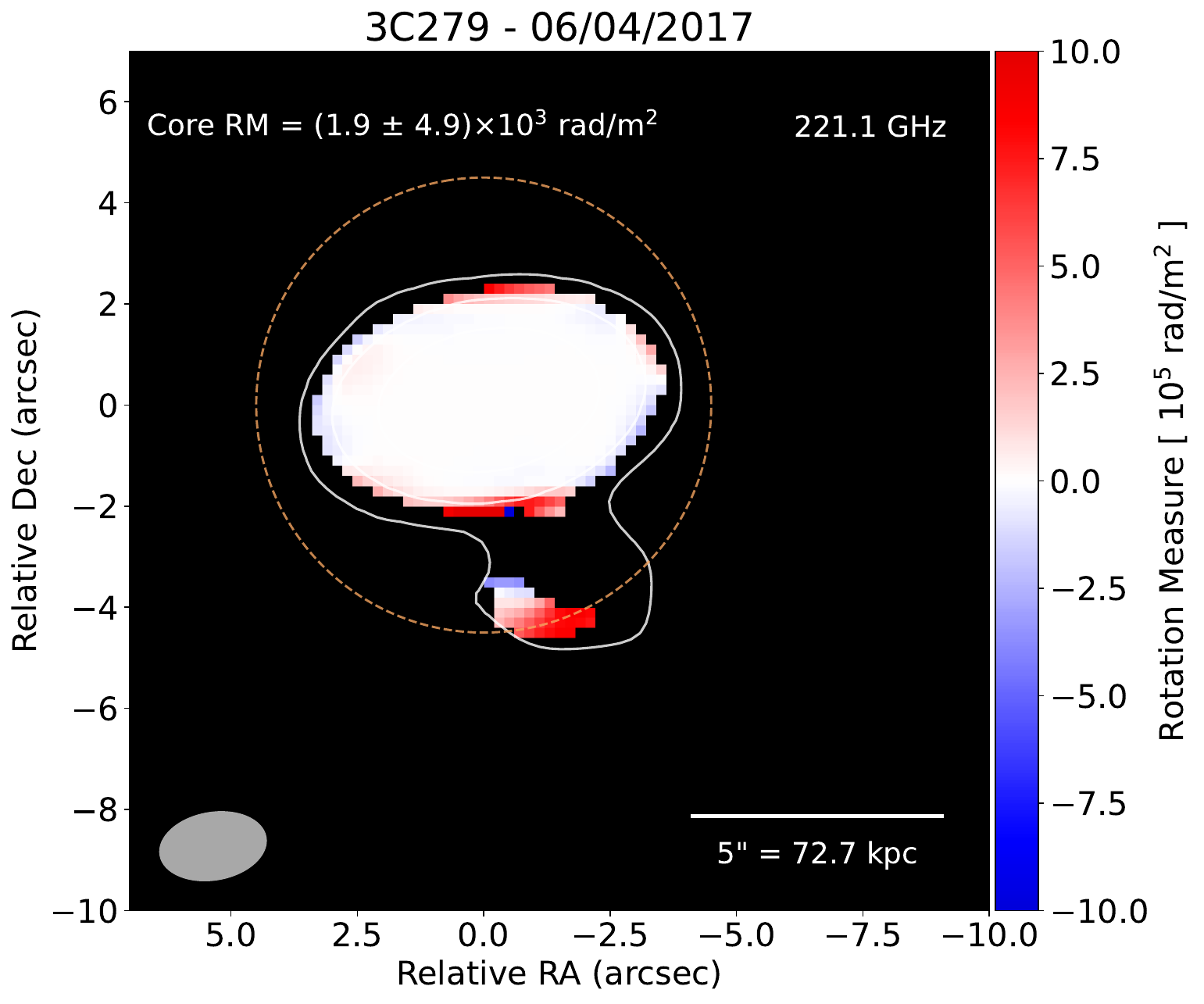}
    \caption{Images of the calibrators 3C273 (top) and 3C279 (bottom) observed in the EHT 2017 campaign. From left to right, Stokes I image, linear polarization image and rotation measure image. Polarization-derived maps also include a dashed circle indicating the inner one-third of the ALMA primary beam, where ALMA guarantees polarization fidelity at the 0.1\% level of Stokes I.}
    \label{fig:4x3_matrix}
\end{figure*}

The \texttt{plots} directory contains pre-rendered maps in \texttt{.png} and
\texttt{.pdf} format for a subset of the observables described in
Sect.\ref{subsection:pol_imaging}, including Stokes~$I$, LP, CP, \texttt{ALPHA}, Depol, and RM.
Each map displays the corresponding quantity in color, overlaid with
Stokes~$I$ contours starting at $4\sigma$.
All plots include ancillary information such as the CLEAN beam,
a reference observing frequency, the core flux density with its associated
uncertainty, and, when applicable, a scale bar converting angular to physical
units. For LP maps, white vectors are overplotted with lengths proportional to
the polarized intensity and orientations corresponding to the EVPA. Polarization-derived maps, i.e., LP, CP, \texttt{ALPHA}, Depol, and RM also contain an overplotted dashed circle, which represents the inner one-third of the primary beam  (see Sect.~\ref{subsection:pol_imaging}). 
As an illustrative example, Fig.~\ref{fig:4x3_matrix} shows Stokes~$I$, LP, and
RM maps of 3C~273 and 3C~279 obtained on 2017 April~6.

\subsubsection{\sgra light curves}

The \texttt{sgra_specifics} folder contains all products derived from the minispiral calibration (Sect.~\ref{subsubsec:sgra}), organized into four subfolders: \texttt{CASA_products}, \texttt{fits}, \texttt{data}, and \texttt{plots}.

The \texttt{data} subfolder stores the light-curve fitting results in \texttt{.dat} files in NumPy array format. These files contain time-resolved flux densities for all four Stokes parameters ($I$, $Q$, $U$, and $V$), their uncertainties as described in Eq.~\ref{eq:fitting}, the observing frequency, and a quality flag (\texttt{GOOD}) marking valid data.

The \texttt{plots} subfolder provides light curves derived from these Stokes parameters, including Stokes~$I$, LP, EVPA, in both \texttt{.png} and \texttt{.pdf} formats. Stokes~$I$ images of \sgra\ produced after minispiral calibration, both per SPW and SPW-combined, are available in \casa\ format in the \texttt{CASA_products} folder and in \texttt{FITS} format in the \texttt{fits} folder.

\section{An overview of science applications}\label{sec:science_case}

The ALMA datasets released through the VAPOLA repository constitute a rich resource
for addressing a wide range of scientific questions across multiple areas of
astrophysics. These full-polarization observations, spanning multiple frequency bands and epochs, enable investigations ranging from
black hole accretion and relativistic jet physics to Galactic structure and the
properties of the interstellar medium. In this section, we outline a representative
selection of science cases that illustrate the breadth, versatility, and scientific
value of the VAPOLA data products.

\subsection{Spectral and polarization studies of AGN}\label{subsec:agn}

High-frequency  observations with ALMA offer a unique view of the innermost regions of AGN, capturing both spectral and polarimetric signatures of synchrotron-emitting plasma near supermassive black holes. Many AGN cores display strong linear polarization, with fractions between 2\% and 15\%, and exhibit high RM  \citep[e.g.,][]{Agudo2018a, Goddi2021, Goddi2025}, reflecting the presence of magnetized plasma and structured fields at the jet base. The detection of steep spectral indices ($\alpha = -1.3$ to $-0.2$) between $3\,$mm and $1.3\,$mm further supports synchrotron-dominated emission.

Polarization encodes valuable information on the geometry and ordering of magnetic fields, but interpreting the intrinsic EVPA requires correcting for Faraday rotation due to magnetized plasma along the line of sight. In the case of an external Faraday screen, the EVPA follows a linear dependence on the square of the wavelength. However, when the emitting and rotating media are co-spatial, the resulting internal Faraday rotation introduces more complex—and often depolarizing—effects \citep{Sullivan2012}.
With its broad frequency coverage and high sensitivity, ALMA observations in VAPOLA enable detailed modeling of the polarization structure of AGN and help constrain the nature of the Faraday screen, particularly when used alongside long-baseline observations from VLBI networks such as the GMVA and the EHT \citep{Goddi2021,eht2021a}. The multi-frequency aspect of the dataset is a key ingredient for this effort, as some models can predict complicated LP and EVPA behaviors across a wide spectral range \citep{Hovatta2019}~,~Carlos et al. (in prep).

Polarimetric monitoring over multiple epochs and bands captures both short- and long-term variability. Variations on timescales of days are typically associated with local changes in the emission region, whether in the inner accretion flow or in a stratified jet \citep{Jorstad2005,MartiVidal2021}. In contrast, slower variations over months to years trace the evolution of extended structures, such as jet precession, changes in magnetic field configuration, or fluctuations in plasma density in the surrounding environment \citep{Jorstad2005,Agudo2018b}.

Therefore, VAPOLA allows for a multi-dimensional characterization of AGN environments, helping to probe magnetic fields, plasma conditions, and the physical coupling between accretion flows and relativistic outflows in the vicinity of supermassive black holes. 

\subsection{Molecular torus in AGN}\label{subsection:torus}

In the unified model of AGN, a compact, dusty, and molecular torus plays a central role in shaping the observed diversity of AGN types. Thanks to ALMA’s angular resolution, it is now possible to directly image and characterize these structures \citep[e.g.,][]{GarciaBurillo2016, GarciaBurillo2024}. Observations of molecular transitions—such as CO, HCN, and HCO$^+$—in the millimeter regime reveal both the morphology and kinematics of the circumnuclear gas \citep[e.g.,][]{Combes2019}.

\begin{figure}[htbp]
  \centering
  \includegraphics[width=\columnwidth]{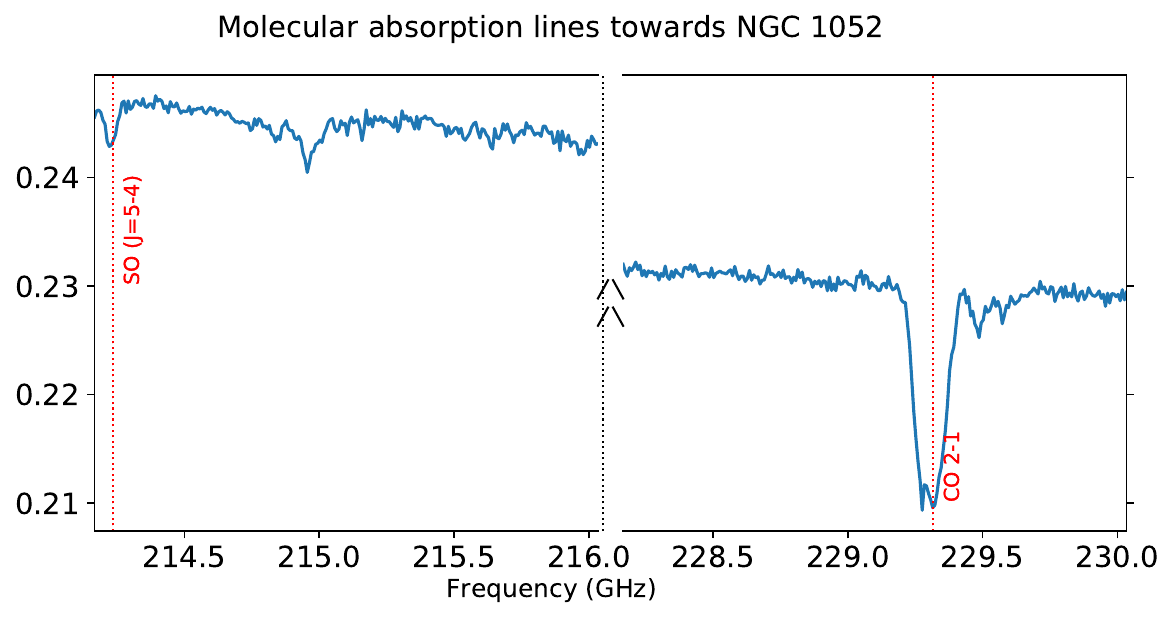}
\caption{Absorption lines previously reported in the literature, detected toward the
molecular torus of NGC~1052 during the 2017 Band~6 observing campaign.
The spectra correspond to SPW~0 (left panel) and SPW~3 (right panel).}
  \label{fig:ngc1052}
\end{figure}

One prominent example is NGC1052, where ALMA has revealed strong molecular absorption features—including CO, HCN, HCO$^+$, SO, SO$_2$, CS, CN, and H$_2$O—against the bright continuum from the nucleus \citep{Kameno2020, Kameno2023}. Several of these lines are also detected in our VAPOLA datasets. For instance, Fig.\ref{fig:ngc1052} (left panel) shows the absorption from SO ($J=5-4$) at 214.357\,GHz (\internaleht{SPW} 0) and CO (2–1) at 230.538\,GHz (\internaleht{SPW} 2) in the  correlation spectra from the EHT 2017 ALMA observations.

\begin{figure}[htbp]
  \centering
  \includegraphics[width=\columnwidth]{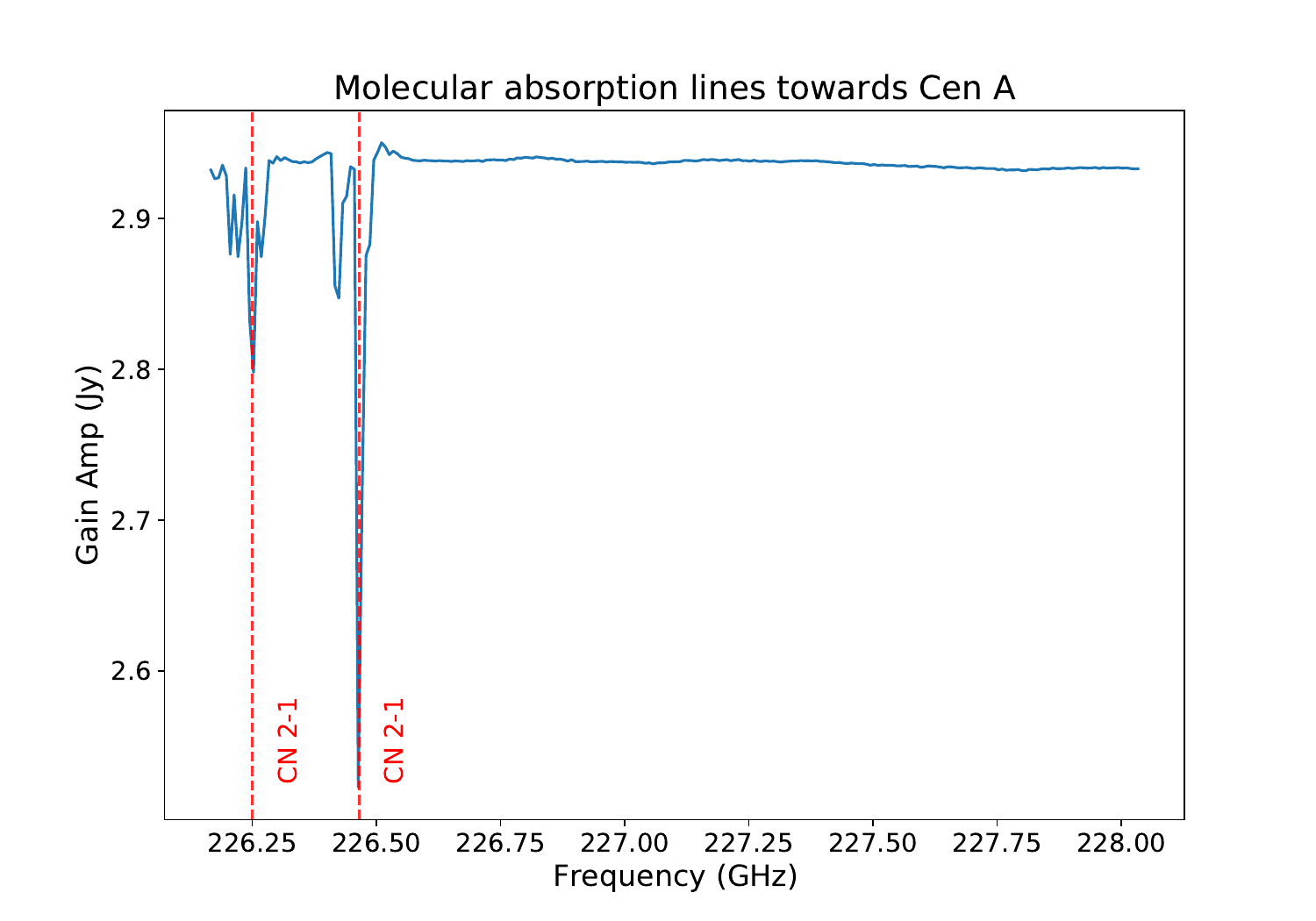}
\caption{Absorption CN lines detected for the first time in these observations toward
Centaurus~A, in  SPW~2 of the Band~6 data obtained on 2017 April~10.}
  \label{fig:absorption_cena}
\end{figure}

Centaurus A (Cen A hereafter) provides another case of circumnuclear molecular gas in a radio galaxy \citep{Espada2017}. Fig.~\ref{fig:absorption_cena}  highlights the detection of absorption lines  potentially associated with the  hyperfine structure of the diatomic molecule CN, with main components near 226.69, 226.89, and 226.90\,GHz. The offsets between the observed line peaks in Fig.~\ref{fig:absorption_cena} and their laboratory rest frequencies can be naturally explained as Doppler shifts caused by gas motion along the line of sight.

More recently, \citet{Boizelle2025} used archival ALMA data, including three VLBI projects, to detect CO(1–0), CO(2–1), and CO(3–2) absorption toward the M87 core. These features were interpreted as arising from optically thin molecular clouds or filaments aligned by chance with the AGN core.

These literature studies show that ALMA enables spatially resolved mapping of the molecular torus, including measurements of its mass, clumpiness, and scale height. In addition, absorption-line studies allow estimates of black hole mass, torus dynamics, and the efficiency of feedback via molecular outflows.
Further observations and analyses of molecular lines with ALMA are essential for testing AGN unification models and for constraining the role of the torus in obscuring and fueling the central engine~\citep[e.g.,][]{GarciaBurillo2016, Combes2019}.

\begin{figure}[htbp]
    \centering
    \includegraphics[width=.5\textwidth]{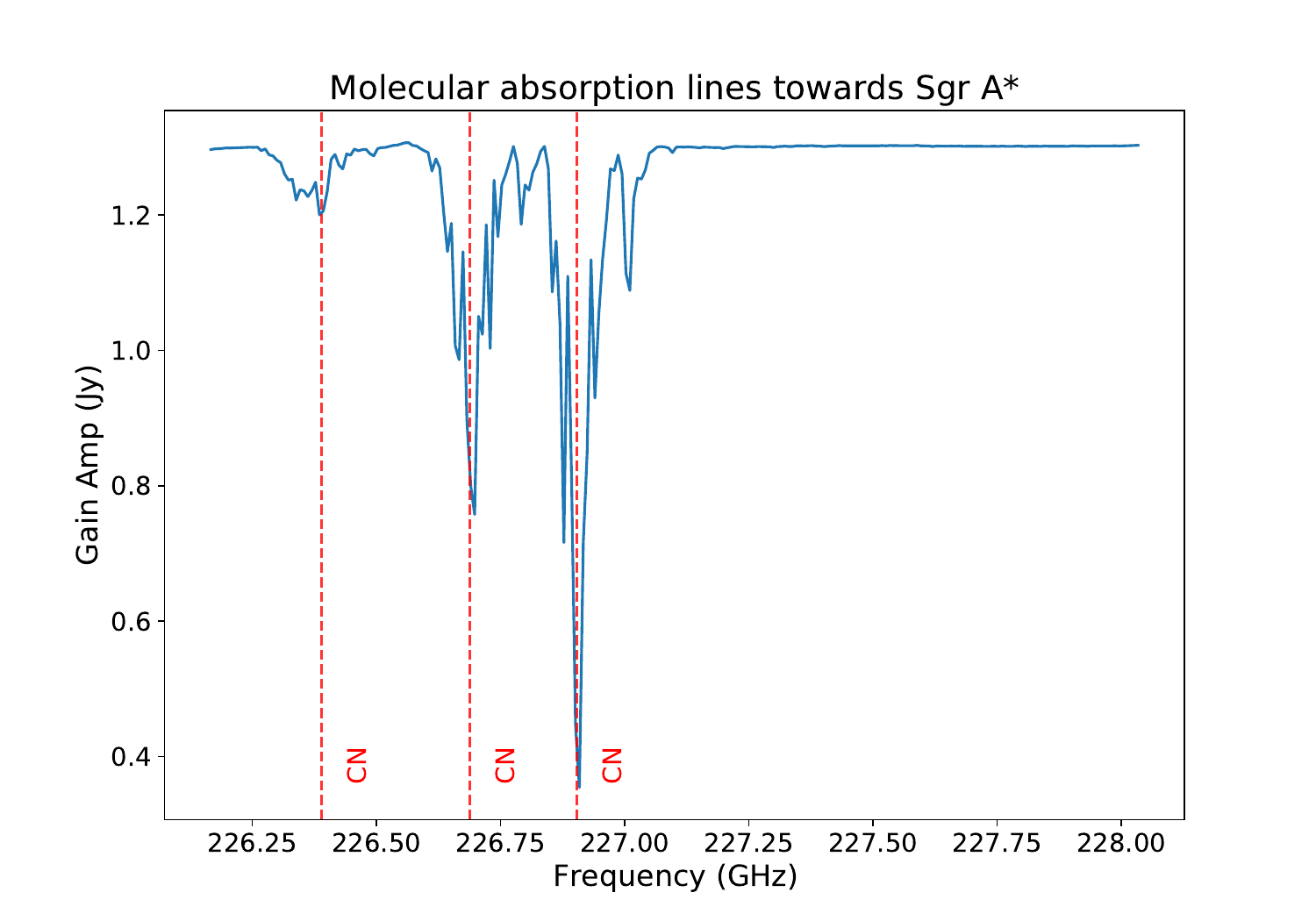}
\caption{Absorption lines toward \sgra observed on 2017 April~6 in SPW~2,
previously reported by \citet{Goddi2021}.}
    \label{fig:absorption_sgra}
\end{figure}

\subsection{Interstellar medium at the Galactic Center}\label{subsection:ismSgra}

Spectra toward strong continuum sources in the Galactic Center frequently reveal
absorption features from interstellar molecules located in the Central Molecular Zone
(CMZ) and in the extended Galactic bulge
\citep[e.g.,][]{Riquelme2018}.
During the 2017 VLBI campaign, \citet{Goddi2021} reported the detection of absorption
features in the continuum spectrum of \sgra, attributed to the hyperfine structure
of the CN molecule near $226.91\,\mathrm{GHz}$
(Fig.~\ref{fig:absorption_sgra}).
Similar CN absorption lines were also observed toward the nearby quasar
J1744--3116, located about $2^\circ$ from \sgra
\citep[see also][]{Riquelme2018}.
Because this source lies outside the densest regions of the CMZ, it probes a more
diffuse interstellar component.
The presence of comparable absorption features along both sightlines suggests either
the existence of a widespread diffuse molecular component across the Galactic Center
region, or the superposition of multiple molecular clouds at different Galactocentric
distances sharing similar chemical properties.

Millimeter observations of molecular tracers such as HCN, CN, HCO$^+$, and CS enable
the decomposition of distinct velocity components and the investigation of
small-scale ISM structures
\citep[e.g.,][]{LisztLucas2001,Tsuboi2018,Henshaw2023}.
ALMA data provided through the VAPOLA repository significantly enhance these studies
by enabling the detection and identification of molecular absorption lines that can
be followed up with dedicated high-resolution spectroscopy.
High spectral resolution observations allow for precise velocity measurements,
constraints on physical conditions (e.g., temperature and density), and insights into
the chemical composition of molecular gas toward the Galactic Center.

While most EHT observations at $\sim230$\,GHz—and more
recently at 345\,GHz—have focused on continuum imaging of black holes and AGN,
spectral-line VLBI remains largely unexplored.
The detection of molecular absorption toward \sgra therefore opens a promising avenue
for using VLBI techniques to localize and characterize cold molecular gas in the
immediate Galactic Center environment.
Such studies provide a novel means to probe the structure, dynamics, and composition
of the CMZ, as well as its potential connection to the central black hole.

\subsection{\sgra time-variability studies}

\sgra is known for its rapid and complex variability linked to accretion and magnetic activity near the event horizon~\citep{Wielgus2022, eht2022a, eht2022b, eht2022c, eht2022d, eht2022e, Albentosa2025, EHTDynamics}. The integration of ALMA in VLBI campaigns allows for high-sensitivity, sub-milliarcsecond resolution observations that can capture these short-term fluctuations \citep{eht2022d}. VAPOLA provides science-ready data sets that directly enable studies such as those of \citep{Wielgus2022} and \cite{Albentosa2025}, where temporal variations in the Stokes parameters have been demonstrated,
with the detection of $Q-U$ loops serving as evidence for an orbiting hotspot in the innermost regions \cite{Wielgus2022b}. Such time-resolved observations are pivotal in discriminating between different flare mechanisms, for instance magnetic reconnection events or turbulent accretion~\citep[e.g.,][]{eht2024b, eht2024c}.

\subsection{Transient searches at high frequencies}

The enhanced sensitivity and resolution provided by ALMA in VLBI mode are equally transformative for transient astronomy.~\cite{Liu2021} and \cite{mus22} performed the first high-frequency (slow) transient searches in the Galactic Center using ALMA data taken during the 2017 Global (GMVA and EHT respectively) campaign. The capability to monitor the sky on relatively short timescales significantly enhances the detection and characterization of transient events to better understand the dynamics of the Galactic Center. These studies target a variety of transient phenomena, including rapid flares in compact objects, fast radio bursts~\citep{Macquart2015} or pulsars~\citep{Torne2023}. VAPOLA offers a multiband multiepoch repository where this phenomena can be explored.

\begin{figure*}[htbp]
  \centering
  \begin{tikzpicture}[>=stealth]

    \node[anchor=south west, inner sep=0] (img1) at (0,0)
    {\includegraphics[width=.4\linewidth]{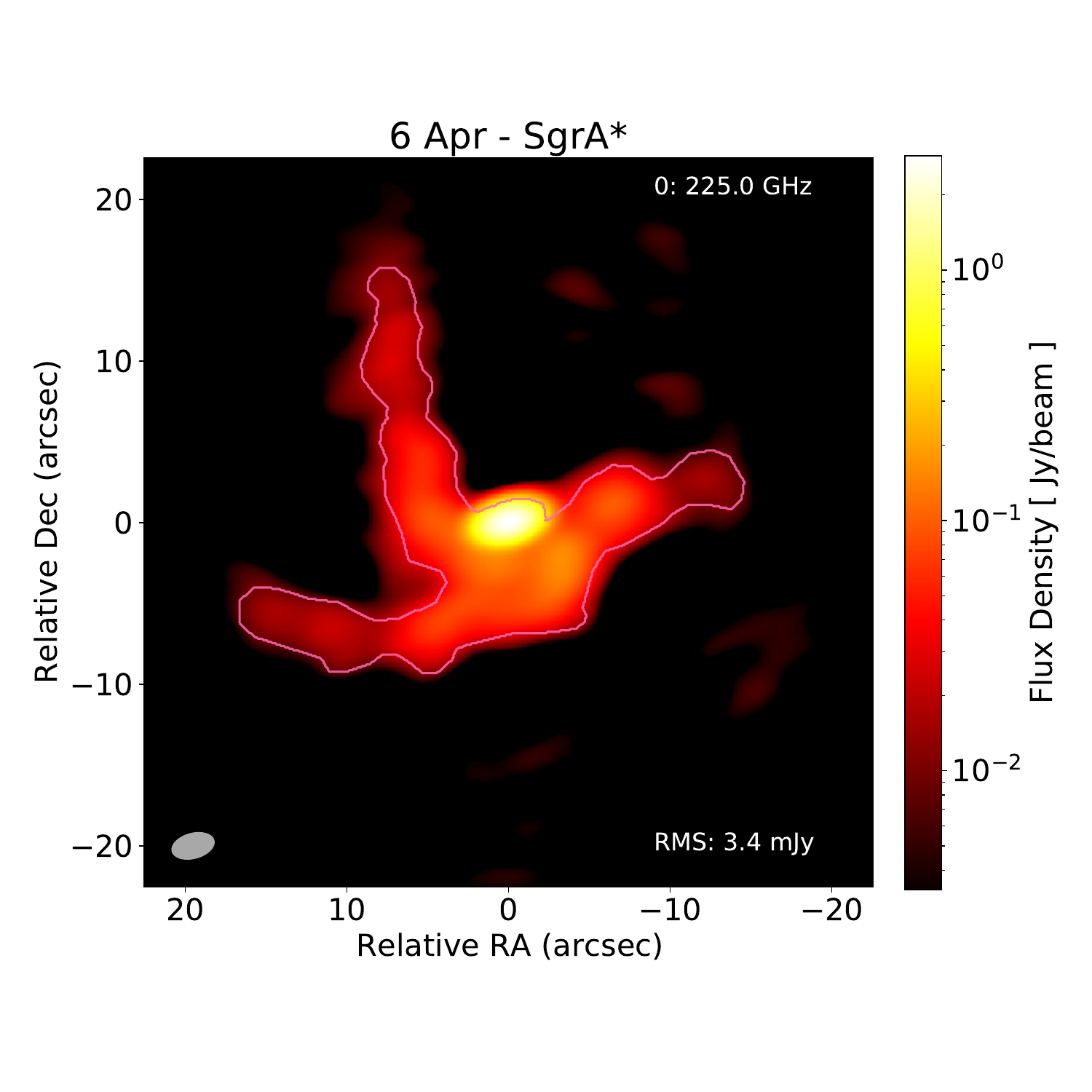}};
    
    \node[anchor=south west, inner sep=0] (img2) at (8,0)
    {\includegraphics[width=.5\linewidth]{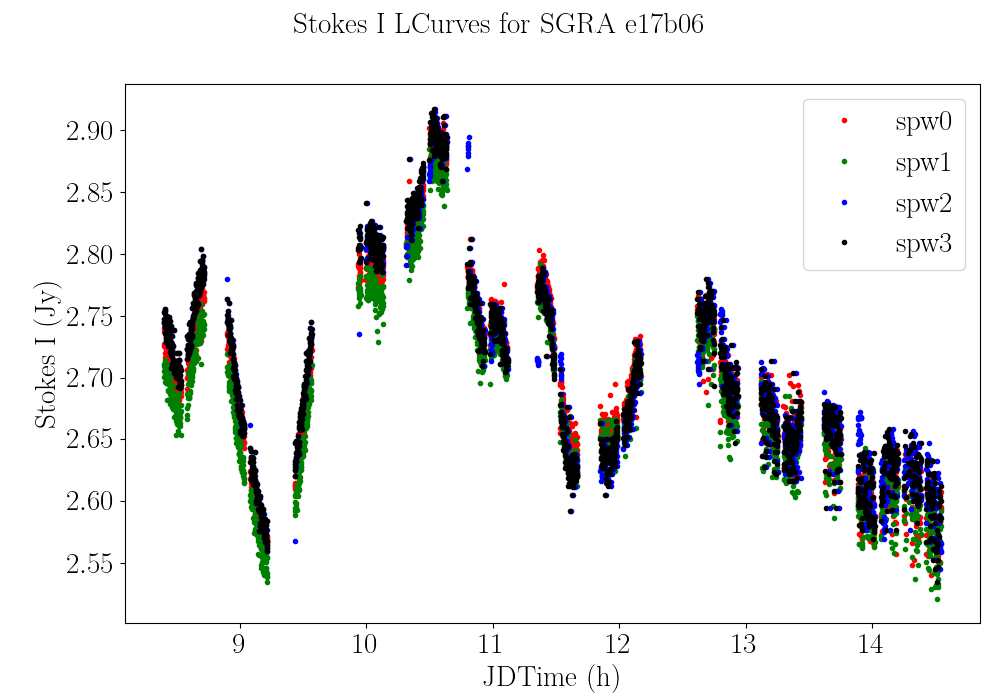}};
    
    \draw[red, thick]
      ([xshift=3.5cm, yshift=4cm] img1.south west) 
      -- ([xshift=1.0cm, yshift=-0.5cm] img2.north west);
    
    \draw[red, thick]
      ([xshift=3.5cm, yshift=3.6cm] img1.south west) 
      -- ([xshift=1.0cm, yshift=.5cm] img2.south west);

  \end{tikzpicture}
  \caption{Left panel: Stokes $I$ image of SgrA* (compact core) and the surrounding minispiral structure (extended emission), observed on 2017 April 6. The synthesized beam is shown as an ellipse in the bottom left corner; the RMS noise level is indicated in the bottom right; the average frequency of the observation is shown in the top right. The white contour outlines the CLEANing mask applied during image reconstruction.
Right panel: Stokes $I$ light curves of SgrA* extracted after applying minispiral calibration on the same date. Different spectral windows (\internaleht{SPW}s) are plotted in different colors.
Both panels show representative science products from the VAPOLA repository.}
\label{fig:sgra_lightcurve}
\end{figure*}

\section{Summary and conclusions}\label{sec:conclusions}

ALMA plays a unique role in millimeter and sub-millimeter astronomy, both as the most
sensitive connected-element interferometer operating at these wavelengths and as a
key element of global VLBI arrays such as the GMVA and the EHT.
Beyond its contribution to VLBI imaging, ALMA observations obtained during VLBI
campaigns provide a wealth of standalone interferometric information, particularly
for polarization, time-domain, and spectral-line studies
\citep{Goddi2019,Goddi2021,Wielgus2022,Wielgus2022b,Albentosa2025,Goddi2025}.
Fully exploiting this potential, however, requires access to consistently calibrated,
science-ready products that go beyond the standard archive deliveries.

In this work we have presented \textbf{VAPOLA}, a public data repository that provides
homogeneous, high-level ALMA products derived from VLBI campaigns conducted since
2017 (Cycle~4), covering Bands~3, 6, and~7.
The repository includes calibrated visibilities, full-Stokes images, and a broad set
of derived polarimetric and spectral diagnostics, delivered in formats optimized for
scientific analysis rather than calibration-oriented workflows.
For intrinsically variable sources such as \sgra, VAPOLA further provides
time-domain interferometric products, including light curves of Stokes parameters and
polarization properties.

All data products are generated through a largely automated processing pipeline that
builds upon the standard QA2 calibration framework.
On the one hand, this approach minimizes user-dependent choices, improves
reproducibility, and enables uniform comparisons across large data sets.
On the other hand, specialized procedures—such as the application of the minispiral calibration for \sgra—allow VAPOLA to address challenges that are not handled by standard archive products.

All non-proprietary VAPOLA data products are publicly available through a dedicated web
portal\footnote{\url{http://vapola.ia2.inaf.it}} and will be updated regularly as new
observations become available.
The repository is designed to evolve alongside improvements in calibration strategies
and software, adopting a transparent versioning scheme that preserves data provenance
and reproducibility.

By providing ready-to-use interferometric polarization data at millimeter wavelengths,
we anticipate that VAPOLA will enable a wide range of investigations, from AGN jet
physics and black hole environments to molecular line studies of AGN molecular tori
and the Galactic interstellar medium, while facilitating broader community access to
ALMA polarization data obtained during VLBI campaigns.

\begin{acknowledgement}

This project was funded by the European Union through NextGenerationEU under RRF M4C2 1.1, project No.~2022YAPMJH; by the Coordena\c{c}~ao de Aperfei\c{c}oamento de Pessoal de N'ivel Superior (CAPES), Brazil, under PROEX grant No.~88887.845378/2023-00; and by the Funda\c{c}~ao de Amparo `a Pesquisa do Estado de S~ao Paulo (FAPESP) under grant No.~2021/01183-8. We acknowledge George Wong for his help and guidance in securing the VAPOLA frontend including the https protocol and logging hashing. 
This paper makes use of the following ALMA data:
\begingroup
\scriptsize
\almaproj{2016.1.00413.V},
\almaproj{2016.1.01114.V},
\almaproj{2016.1.01116.V},
\almaproj{2016.1.01154.V},
\almaproj{2016.1.01176.V},
\almaproj{2016.1.01198.V},
\almaproj{2016.1.01216.V},
\almaproj{2016.1.01290.V},
\almaproj{2016.1.01404.V},
\almaproj{2017.1.00795.V},
\almaproj{2017.1.00797.V},
\almaproj{2017.1.00841.V},
\almaproj{2017.1.00842.V},
\almaproj{2017.1.00985.V},
\almaproj{2017.1.00991.V},
\almaproj{2017.1.01339.V},
\almaproj{2017.1.01514.V},
\almaproj{2019.1.00183.V},
\almaproj{2019.1.00636.V},
\almaproj{2019.1.01218.V},
\almaproj{2019.1.01797.V},
\almaproj{2019.1.01812.V},
\almaproj{2021.1.00767.V},
\almaproj{2021.1.00889.V},
\almaproj{2021.1.00906.V},
\almaproj{2021.1.00910.V},
\almaproj{2021.1.01156.V},
\almaproj{2021.1.01324.V},
\almaproj{2021.1.01431.V},
\almaproj{2021.1.01458.V},
\almaproj{2021.1.01515.V},
and \almaproj{2022.1.01268.V}.
\endgroup
ALMA is a partnership of ESO (representing its member states),
NSF (USA) and NINS (Japan), together with NRC (Canada), NSTC and
ASIAA (Taiwan), and KASI (Republic of Korea), in cooperation with
the Republic of Chile. The Joint ALMA Observatory is operated by
ESO, AUI/NRAO and NAOJ.
\\
\textbf{Software Availability}\label{sec:softwareAvailability}
\\
\url{https://casa.nrao.edu/}
\url{https://mural.uv.es/imarvi/docums/uvmultifit/}
\end{acknowledgement}

\bibliographystyle{aa}
\bibliography{lib}{}

\begin{appendix}

\onecolumn
\section{Additional tables}

\begin{table*}[htpb]
  \centering
  \setlength{\tabcolsep}{4pt}
  \small
  \caption{Example of the observation summary file of ALMA tracks \internaleht{contained} in VAPOLA repository.}

\newcolumntype{P}[1]{>{\raggedright\arraybackslash}p{#1}} 
\newcolumntype{C}[1]{>{\centering\arraybackslash}p{#1}}   

\begin{tabularx}{\textwidth}{
  P{2.4cm} C{1.5cm} C{1cm} C{1.cm} C{1.cm} C{1.cm} C{1.cm}
  P{2.4cm} P{1.5cm} C{1.cm} C{1.cm}
}
\toprule
\multirow{2}{*}{\textbf{Track}} &
\multirow{2}{*}{\textbf{Date}} &
\multirow{2}{*}{\makecell{\textbf{Nr.}\\\textbf{ant.}}} &
\multicolumn{2}{c}{\makecell[b]{\textbf{Baseline}\\\textbf{length [m]}}} &
\multicolumn{2}{c}{\makecell[b]{\textbf{UT [h]}}} &
\multirow{2}{*}{\textbf{Project}} &
\multirow{2}{*}{\textbf{Source}} &
\multirow{2}{*}{\makecell{\textbf{Obs.}\\\textbf{mode}}} &
\multirow{2}{*}{\makecell{\textbf{Int.}\\\textbf{time [min]}}}
\\
\cmidrule(lr){4-5}\cmidrule(lr){6-7}
& & & \textbf{min} & \textbf{max} & \textbf{start} & \textbf{end} & & & & \\
\midrule

\multicolumn{11}{c}{\textbf{\underline{Band B3}}}\\
\multirow{4}{*}{2016.1.00413.V}
& \multirow{4}{*}{2017-04-05}
& \multirow{4}{*}{37}
& \multirow{4}{*}{15.06}
& \multirow{4}{*}{278.86}
& \multirow{4}{*}{22:11}
& \multirow{4}{*}{09:13}
& \multirow{4}{*}{\makecell{2016.1.00413.V}}\\
& & & & & & & & 4C09.57 & V & 8.06\\
& & & & & & & & NRAO530 & V & 28.76\\
& & & & & & & & \sgra & V & 173.64\\
& & & & & & & & J1924-2914 & V & 8.06\\

\multicolumn{11}{c}{\textbf{\ldots}}\\
\midrule
\midrule

\multicolumn{11}{c}{\textbf{\underline{Band B6}}}\\
\multirow{4}{*}{e17d05}
& \multirow{4}{*}{2017-04-05}
& \multirow{4}{*}{37}
& \multirow{4}{*}{15.06}
& \multirow{4}{*}{278.86}
& \multirow{4}{*}{22:11}
& \multirow{4}{*}{09:13}
& \multirow{4}{*}{\makecell{2016.1.01176.V,\\ 2016.1.01154.V}} \\
& & & & & & & & 3C279 & V & 79.03\\
& & & & & & & & OJ287 & V & 93.00\\
& & & & & & & & 4C 01.28 & V & 15.32\\
& & & & & & & & M87 & V & 102.95\\
\addlinespace
\cdashline{8-11}[1pt/2pt]
\addlinespace
& & & & & & & 2016.1.01176.V \\
& & & & & & & & 3C279 & V & 79.03\\
& & & & & & & & M87 & V & 102.95\\
& & & & & & & &  3C279 & NV & 12.10\\
& & & & & & & & 4C01.28 & NV & 29.23\\
& & & & & & & & J0750+1231 & NV & 2.55\\
& & & & & & & & J0837+2454 & NV & 9.68\\
& & & & & & & & 3C275.1 & NV & 7.53\\
& & & & & & & & J1246-0730 & NV & 3.76\\

\multicolumn{11}{c}{\textbf{\ldots}}\\
\midrule
\midrule

\multicolumn{11}{c}{\textbf{\underline{Band B7}}}\\

\multirow{4}{*}{e21a14}
& \multirow{4}{*}{2021-04-19}
& \multirow{4}{*}{28-30}
& \multirow{4}{*}{15.06}
& \multirow{4}{*}{650.31}
& \multirow{4}{*}{01:12}
& \multirow{4}{*}{06:08}
& \multirow{4}{*}{2019.1.01812.V} \\
& & & & & & & & \sgra & V & 187.62\\
& & & & & & & & J1924-2914 & V & 25.27\\
& & & & & & & & PKS1741-03 & V & 22.28\\
& & & & & & & & \sgra & NV & 13.44\\
& & & & & & & & J1924-291 & NV & 28.22\\

\midrule

\multirow{4}{*}{e23d15}
& \multirow{4}{*}{2023-04-15}
& \multirow{4}{*}{33}
& \multirow{4}{*}{15.28}
& \multirow{4}{*}{650.30}
& \multirow{4}{*}{02:20}
& \multirow{4}{*}{14:10}
& \multirow{4}{*}{\makecell{2022.1.01268.V}} \\
& & & & & & & & \sgra & V & 252.40\\
& & & & & & & & 3C279 & V & 30.71\\
& & & & & & & & 3C273 & V & 11.29\\
& & & & & & & & PKS1335-127 & V & 20.70\\
& & & & & & & & J1924-291 & V & 37.36\\
& & & & & & & & \sgra & NV & 46.37\\
& & & & & & & & 3C279 & NV & 10.08\\
& & & & & & & & J1924-291 & NV & 56.99\\

\multicolumn{11}{c}{\textbf{\ldots}}\\

\midrule
\end{tabularx}
\tablefoot{
Rows are grouped by observing band, indicated at the top of each block. 
Dates are given in the format YYYY–MM–DD. 
The ``Nr. ant.'' column reports the range (min–max) of phased antennas during the track; if a single number is listed, the number of antennas remained constant throughout. 
``Baseline length'' gives the minimum and maximum projected baselines [m]. 
``UT start/end'' indicate the start and end times of the track observations. 
``Obs. mode'' distinguishes between VLBI-mode (V) and non-VLBI-mode (NV). 
Dots indicate additional tracks omitted here for space. 
``Int. time'' is the net on-source integration time per track after calibration and flagging. 
For a given track and source, this value represents the cumulative on-source time within that track. 
The complete and regularly updated version of this table is available on the VAPOLA website at \url{http://vapola.ia2.inaf.it/download.html} (\texttt{Tables/observations\_summary/tracks\_summary.csv}).
}
\label{tab:list_projects}

\end{table*}

\twocolumn
\section{Isolation Forest}
\label{app:isolation_forest}

In this appendix, we briefly describe the Isolation Forest (IF) flagging procedure used in Sect.~\ref{subsection:flagging}. The algorithm, introduced by \citet{IsolationForest}, is an unsupervised anomaly detection method designed to identify data points that are relatively isolated with respect to the bulk of the dataset. This technique and related anomaly-detection approaches have been tested and explored in astronomy in a range of applications, including time-domain analyses and large survey datasets (e.g., \citealt{Ishida2019,Lochner2021}). IF is based on the idea that anomalous points are ``few and different'', and can be separated from the main distribution through recursive random partitioning. In practice, the algorithm constructs an ensemble of binary trees by randomly selecting subsets of the data and partitioning the feature space. Points that require fewer splits to become isolated are assigned higher anomaly scores.

In our implementation, the algorithm is applied to the complex visibilities, treating the real and imaginary components jointly. This allows the identification of outliers in the full complex plane, rather than relying on independent thresholds in amplitude or phase.
We stress that the IF procedure is used in a conservative way within the pipeline. It is applied only after QA2 calibration and flagging, and only to compact sources lacking significant extended emission.

By default, the contamination parameter is set to a maximum value of 5\%, which represents an upper bound on the fraction of additional visibilities that may be flagged after QA2 processing. In practice, the fraction of flagged data is typically smaller.
We note that the algorithm identifies statistically isolated points in the visibility distribution and therefore does not explicitly account for astrophysical or spectral structures. As a result, coherent features such as absorption lines are not treated as anomalous, which is consistent with the intended use of the method.

\begin{figure*}
    \centering

    \begin{subfigure}{\linewidth}
        \centering
        \includegraphics[width=.49\linewidth]{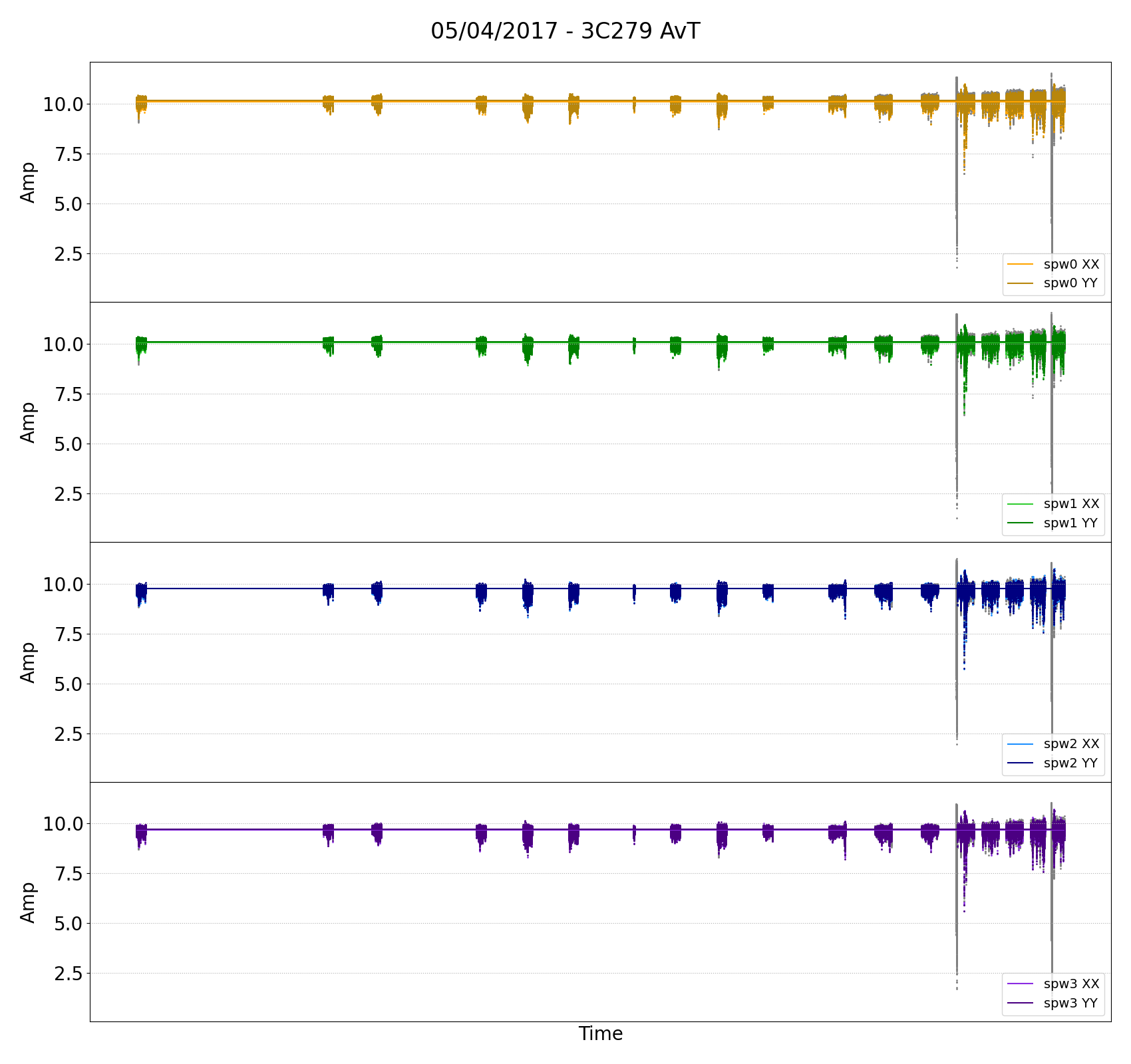}
        \includegraphics[width=.49\linewidth]{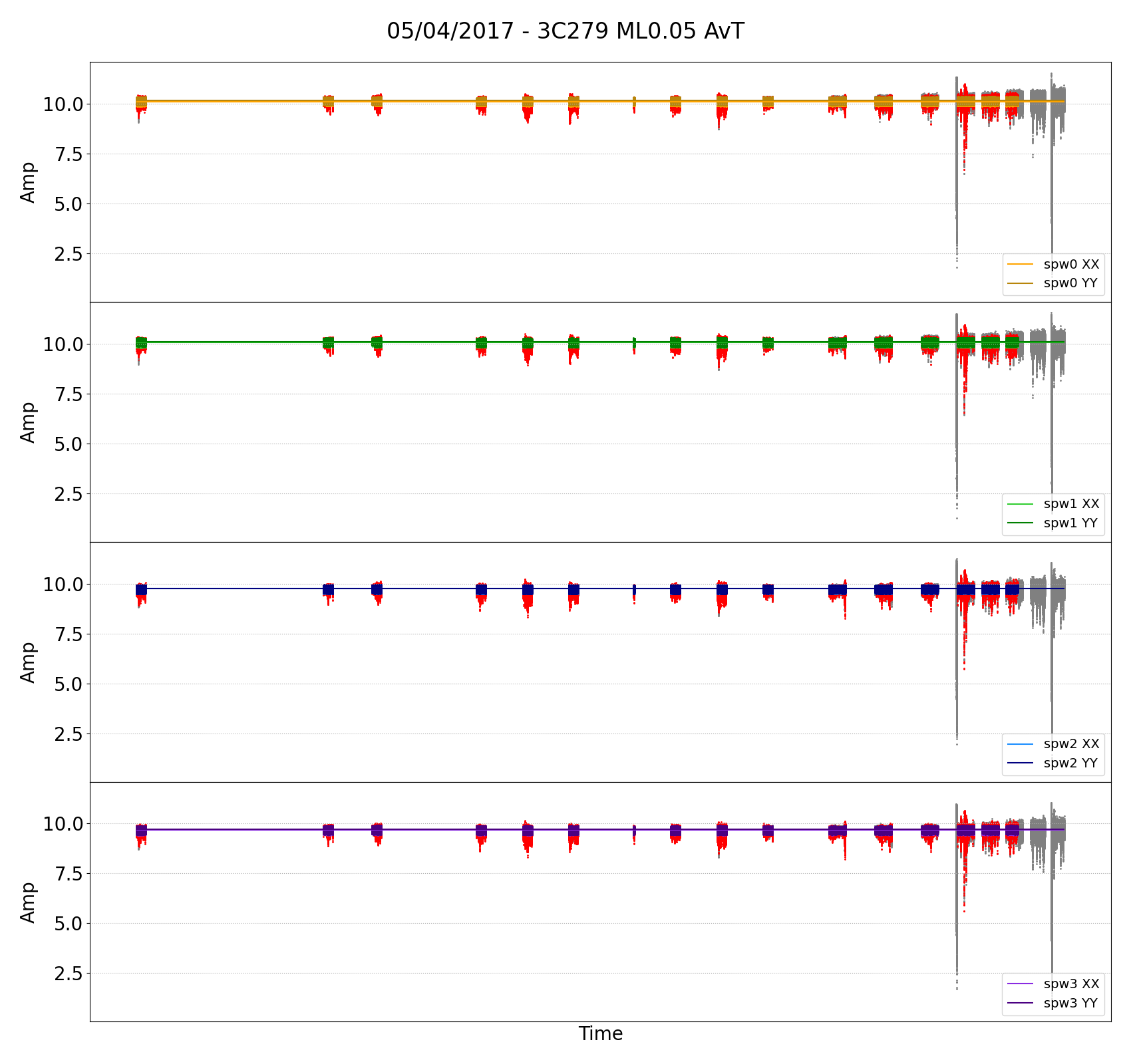}
    \end{subfigure}

    \vspace{0.5em}

    \begin{subfigure}{\linewidth}
        \centering
        \includegraphics[width=.49\linewidth]{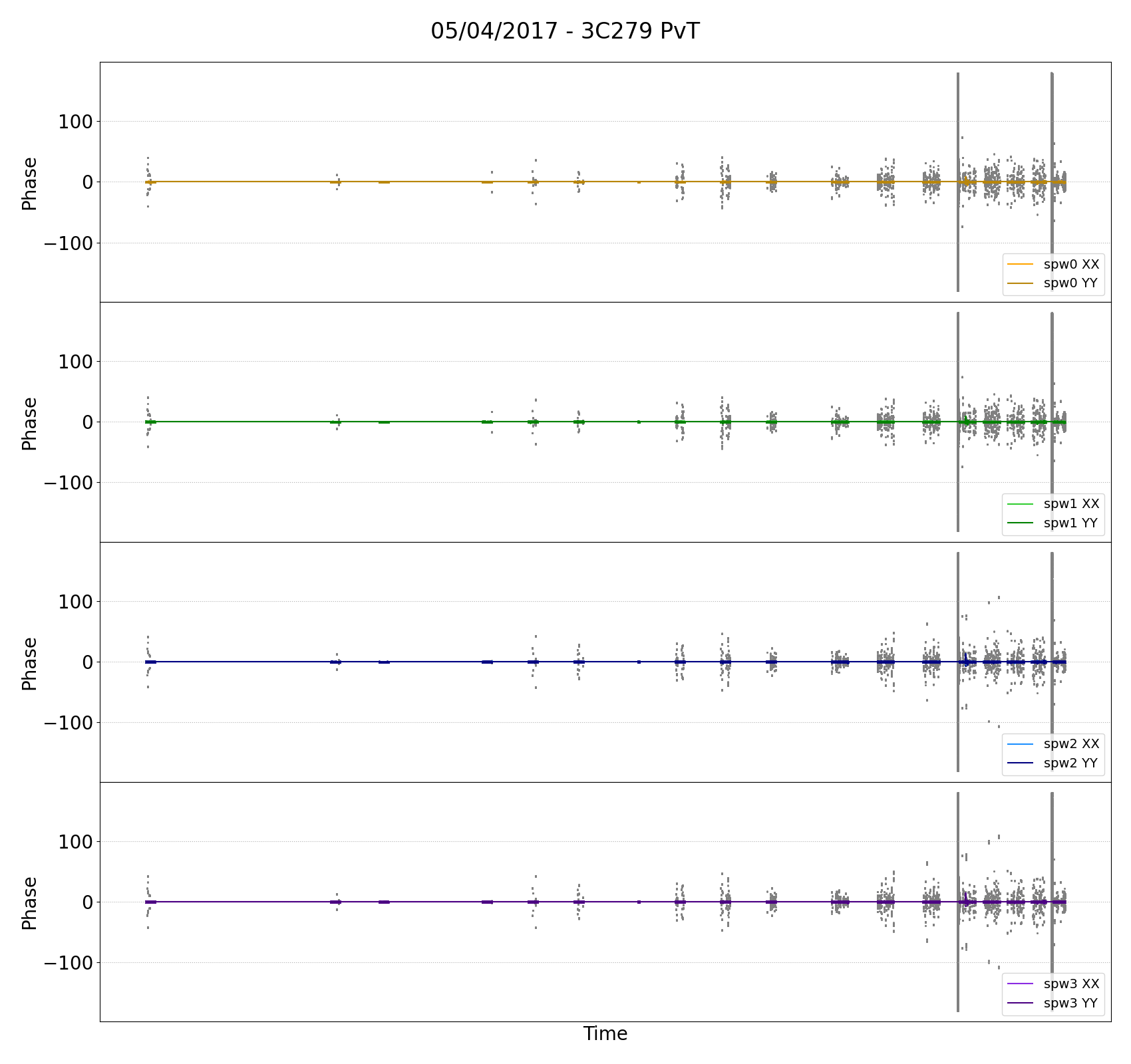}
        \includegraphics[width=.49\linewidth]{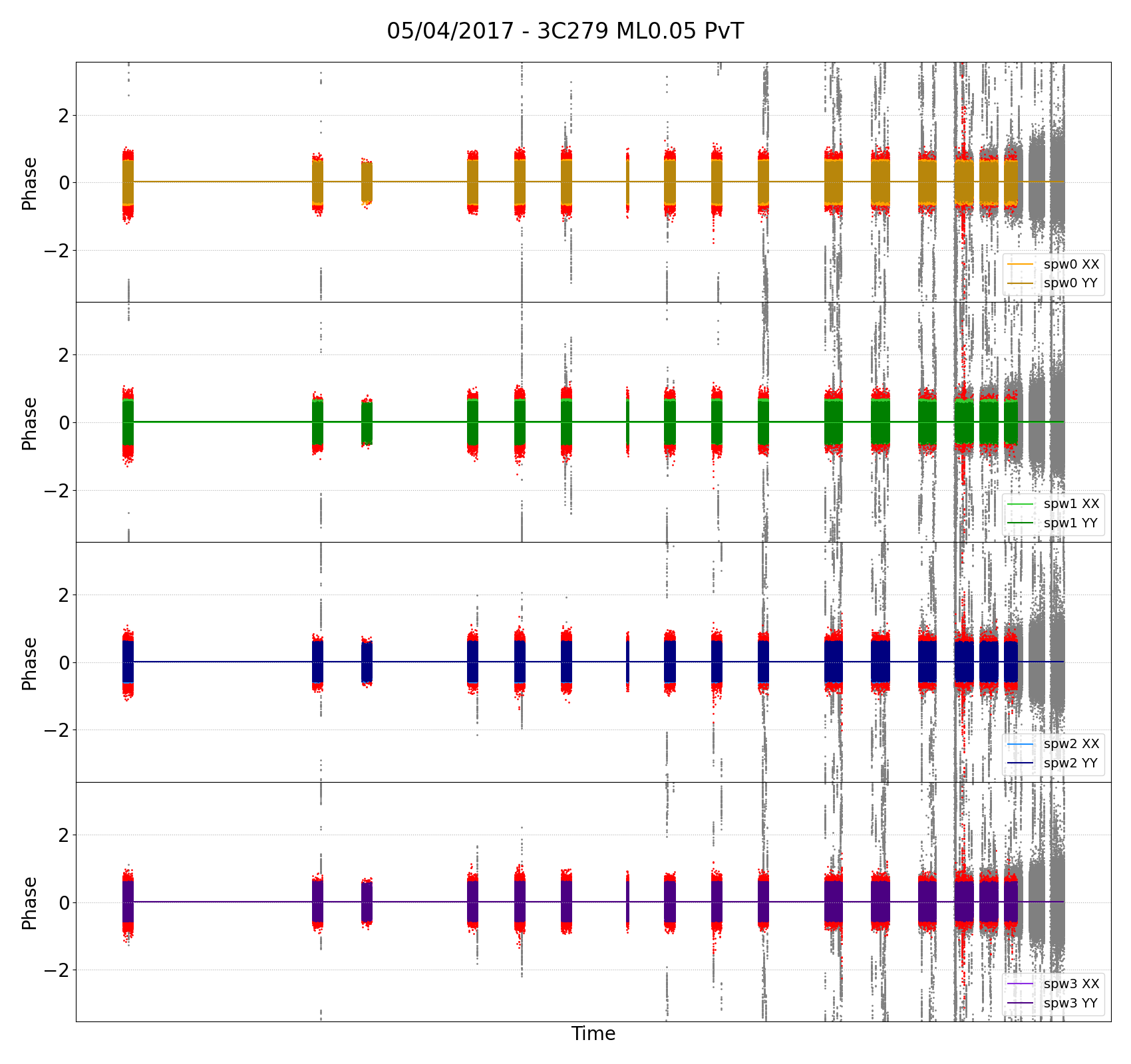}
    \end{subfigure}

    \caption{Amplitude (top) and phase (bottom) versus time of 3C279 during the observation carried out on April 5, 2017. 
    Left panels show the initial state with only QA2 flags. Right panels show the data after IF. The gray points mark the data flagged during QA2 while the red points show those points flagged by IF. }Diagnostic plots can be found in the VAPOLA webpage.
    \label{fig:flag_diag}
\end{figure*}

An example of the flagging process is shown in Fig.~\ref{fig:flag_diag}, using the Band 6 observations of 3C279 obtained on 2017 April 5. The left column displays the visibilities flagged during the standard QA2 calibration (gray points), while the right column highlights the additional outliers identified by the IF algorithm (red points). The figure illustrates that the IF procedure removes only a small number of isolated outliers,  thereby complementing the more extensive QA2 flagging applied during the calibration stage.

\section{VAPOLA webpage}\label{app:vapola_download}

The VAPOLA project is hosted at the Italian Center for Astronomical Archives
(IA2, Trieste, Italy), which provides an Infrastructure-as-a-Service (IaaS) solution.
This consists of a dedicated virtual machine (VM) equipped with 20~TB of fast online
storage.
IA2 is responsible for the physical hardware and virtualization layer, while the
VAPOLA collaboration retains full administrative control over the guest operating
system, runtime environment, and all application layers. 
To accommodate future growth of the repository, the infrastructure supports the
migration of older or less frequently accessed datasets from online storage to a
long-term cold storage system, leveraging the tape library facilities managed by IA2.

In this Appendix we describe the main components of the VAPOLA webpage and its user
interaction model. Additional information and usage examples are available directly
on the website.

\vspace{0.3cm}
Each page of the VAPOLA website displays, in the upper-left corner, a navigation menu
(the ``hamburger'' icon, shown as three horizontal bars), which allows users to switch
between the following sections:
\begin{itemize}
    \item Home
    \item Download
    \item Documentation
\end{itemize}
The first two sections are described in detail below.
The \emph{Documentation} page provides links to external resources, user guides, and
related material.

\subsection{Home page}

The \emph{Home} (landing) page presents an overview of the VAPOLA project, its scientific
scope, and its objectives.
It describes the nature and organization of the repository, the types of data products
provided, and how the data are structured within the archive.
PIs are reminded to authenticate in order to access any
proprietary datasets.
The page also lists contact information for the main contributors to the project.

\subsection{Download page}

The VAPOLA download section, accessible at
\url{https://vapola.ia2.inaf.it/download.html},
provides a web-based client interface that allows users to browse, select, and
retrieve both observational-level and advanced data products.
The webpage acts as a frontend client communicating with a dedicated backend service
responsible for data access, filtering, and packaging.

The \emph{Download} page provides step-by-step, self-explanatory instructions for
accessing and retrieving data products.
Users may authenticate by clicking the login icon in the upper-right corner of the
page.
Depending on the user role, the informational message displayed in the right-hand panel
is updated dynamically, indicating which datasets are visible and accessible under the
permissions associated with the current role.

Users can download individual files directly by clicking the \emph{Download} icon
associated with each table row.
For collection-level data retrieval, two additional download modalities are supported:

\paragraph{Bulk selection.}
Checkboxes next to each file allow users to select multiple entries.
The accumulated list of selected items is displayed via the star-shaped
\emph{wishlist} icon located in the upper-right corner of the page.
Clicking this icon expands the full list of selected files, from which individual items
may also be removed.
The \emph{Download Selected} button triggers the creation and download of a compressed
\texttt{.tar.gz} archive.
A pop-up window provides additional information about the download process.

\paragraph{Product-level filtering.}
By combining facet-based filters with checkbox selection, users can assemble customized
download bundles spanning multiple sources and data types (e.g.\ all full-polarization
maps in Band~3 acquired within a given time range).
These bundles can be retrieved either through a single-click repository download or via
an automatically generated \texttt{.sh} shell script.
Filtering is supported using several criteria: observing mode (VLBI or non-VLBI),
product type (as described in Sect.~\ref{sec:architecture}), and source name.
While the observing-mode filter allows only a single selection, both the product-type
and source filters support multiple selections.

For each selected source, a color-coded container is displayed.
This container allows further refinement by observing year, frequency band, and
observing date.
Filtering is hierarchical: only frequency bands observed in a given year are shown,
and only observing dates corresponding to the selected year and band are displayed.
Within this structure, users may independently select multiple years, bands, and days
for each source.

In both bulk-selection and product-level filtering modes, the resulting compressed
archives additionally include the \texttt{.bib} file associated with this work, enabling
users to reference the repository directly.

All data transfers are performed over HTTPS.
Progress indicators and status messages guide users through multi-file download
operations, while built-in rate limiting prevents excessive load on the backend service
during periods of high demand.

\section{Data versioning and releases}
\label{data_releases}
As described in Sect.~\ref{ScriptVersioning}, updates to the QA2 calibration script suite periodically introduce improved procedures and new functionalities (e.g., refined cross-hand delay and flux calibrations). Whenever such updates affect the calibration process, we plan to reprocess the relevant datasets and issue updated VAPOLA data releases that supersede previous ones. These reprocessings may include revised choices of flux, bandpass, or polarization calibrators, as well as modified flagging strategies.

To ensure full reproducibility, detailed documentation accompanies each data release. This includes:
\begin{itemize}
    \item flag tables and calibration metadata;
    \item lists of calibrators used for flux, bandpass, and polarization;
    \item numerical values adopted in calibration (e.g., flux density and spectral index of the flux calibrator);
    \item information on whether polarization calibration models were derived from VLBI or non-VLBI scans.
\end{itemize}

All documentation is available in the VAPOLA data repository alongside the corresponding dataset.

\vspace{0.3cm}
\noindent
The VAPOLA archive follows a data versioning strategy designed to guarantee traceability of any modification while maintaining efficient use of storage resources. Only the most recent version of each dataset is stored in the primary online repository, whereas previous versions are securely preserved in a long-term cold-storage system managed by IA2. This approach minimizes redundancy in the main archive while ensuring that all past releases remain recoverable upon request.

\vspace{0.3cm}
\noindent
Each modification that affects the data---for instance, an updated flux calibration or a replacement of observational metadata---triggers the creation of a new version. A dedicated log table, accessible from the VAPOLA webpage, records all versioning events. For each dataset, the log includes:
\begin{itemize}
    \item the dataset identifier and data path;
    \item version number (incremental from~1~to~$N$);
    \item date of modification;
    \item brief description of the changes and rationale.
\end{itemize}
This mechanism ensures full transparency of the data provenance and revision history.

\vspace{0.3cm}
\noindent
Although users are encouraged to use the latest available version, earlier releases remain accessible upon justified request. Instructions for retrieval are provided on the same webpage.

\end{appendix}

\end{document}